\documentclass[a4paper,12pt]{article}
\pdfoutput=1
\usepackage{epsfig}
\usepackage{amssymb}
\usepackage{amsfonts}
\usepackage{amsmath}
\usepackage{euscript}
\usepackage{verbatim}
\usepackage{latexsym}
\usepackage{graphicx}
\usepackage{caption}
\usepackage{float}
\usepackage{subcaption}
	\usepackage[T1]{fontenc}

\newskip\humongous \humongous=0pt plus 1000pt minus 1000pt

\newif\ifdtup

\catcode`\@=11

\@addtoreset{equation}{section}

\def\@normalsize{\@setsize\normalsize{15pt}\xiipt\@xiipt
\abovedisplayskip 14pt plus3pt minus3pt%
\belowdisplayskip \abovedisplayskip
\abovedisplayshortskip \z@ plus3pt%
\belowdisplayshortskip 7pt plus3.5pt minus0pt}

\def\small{\@setsize\small{13.6pt}\xipt\@xipt
\abovedisplayskip 13pt plus3pt minus3pt%
\belowdisplayskip \abovedisplayskip
\abovedisplayshortskip \z@ plus3pt%
\belowdisplayshortskip 7pt plus3.5pt minus0pt
\def\@listi{\parsep 4.5pt plus 2pt minus 1pt
     \itemsep \parsep
     \topsep 9pt plus 3pt minus 3pt}}

\relax

\catcode`@=12

\catcode`\@=11

\def\section{\@startsection{section}{1}{\z@}{3.5ex plus 1ex minus
   .2ex}{2.3ex plus .2ex}{\large\bf}}

\def\SymBoxes#1#2#3#4{\newdimen\un@t \un@t#3%
\raisebox{#1}{\rule{#2\un@t}{#4}\hskip-#2\un@t
\@tempdimb\un@t \advance\@tempdimb by-#4\@tempcntb#2\relax%
\@whilenum{\@tempcntb>0}\do{
\rule{#4}{\un@t}\hskip\@tempdimb \advance\@tempcntb by\m@ne}%
\hskip-#2\un@t \rule[\un@t]{#2\un@t}{#4}%
\rule[\un@t]{#4}{#4}\hskip-#4
\rule{#4}{\un@t}}\hskip-#4}                

\begin{document}

\newcommand{\beq}{\begin{equation}}
\newcommand{\eeq}{\end{equation}}
\newcommand{\bea}{\begin{eqnarray}}
\newcommand{\eea}{\end{eqnarray}}
\newcommand{\beas}{\begin{eqnarray*}}
\newcommand{\eeas}{\end{eqnarray*}}
\newcommand{\defi}{\stackrel{\rm def}{=}}
\newcommand{\non}{\nonumber}
\newcommand{\bquo}{\begin{quote}}
\newcommand{\enqu}{\end{quote}}
\renewcommand{\(}{\begin{equation}}
\renewcommand{\)}{\end{equation}}
\def \eqn#1#2{\begin{equation}#2\label{#1}\end{equation}}

\def\e{\epsilon}
\def\IZ{{\mathbb Z}}
\def\IR{{\mathbb R}}
\def\IC{{\mathbb C}}
\def\IQ{{\mathbb Q}}
\def\de{\partial}
\def\Tr{ \hbox{\rm Tr}}
\def\H{ \hbox{\rm H}}
\def\HE{ \hbox{$\rm H^{even}$}}
\def\HO{ \hbox{$\rm H^{odd}$}}
\def\K{ \hbox{\rm K}}
\def\Im{ \hbox{\rm Im}}
\def\Ker{ \hbox{\rm Ker}}
\def\const{\hbox {\rm const.}}
\def\o{\over}
\def\im{\hbox{\rm Im}}
\def\re{\hbox{\rm Re}}
\def\bra{\langle}\def\ket{\rangle}
\def\Arg{\hbox {\rm Arg}}
\def\Re{\hbox {\rm Re}}
\def\Im{\hbox {\rm Im}}
\def\exo{\hbox {\rm exp}}
\def\diag{\hbox{\rm diag}}
\def\longvert{{\rule[-2mm]{0.1mm}{7mm}}\,}
\def\a{\alpha}
\def\dag{{}^{\dagger}}
\def\tq{{\widetilde q}}
\def\p{{}^{\prime}}
\def\W{W}
\def\N{{\cal N}}
\def\hsp{,\hspace{.7cm}}

\def\br{\nonumber}
\def\IZ{{\mathbb Z}}
\def\IR{{\mathbb R}}
\def\IC{{\mathbb C}}
\def\IQ{{\mathbb Q}}
\def\IP{{\mathbb P}}
\def \eqn#1#2{\begin{equation}#2\label{#1}\end{equation}}

\newcommand{\C}{\ensuremath{\mathbb C}}
\newcommand{\Z}{\ensuremath{\mathbb Z}}
\newcommand{\R}{\ensuremath{\mathbb R}}
\newcommand{\rp}{\ensuremath{\mathbb {RP}}}
\newcommand{\cp}{\ensuremath{\mathbb {CP}}}
\newcommand{\vac}{\ensuremath{|0\rangle}}
\newcommand{\vact}{\ensuremath{|00\rangle}                    }
\newcommand{\oc}{\ensuremath{\overline{c}}}
\newcommand{\psizero}{\psi_{0}}
\newcommand{\phizero}{\phi_{0}}
\newcommand{\hzero}{h_{0}}
\newcommand{\psiin}{\psi_{\rh}}
\newcommand{\phiin}{\phi_{\rh}}
\newcommand{\hin}{h_{\rh}}
\newcommand{\rh}{r_{h}}
\newcommand{\rb}{r_{b}}
\newcommand{\psibnd}{\psi_{0}^{b}}
\newcommand{\psibndp}{\psi_{1}^{b}}
\newcommand{\phibnd}{\phi_{0}^{b}}
\newcommand{\phibndp}{\phi_{1}^{b}}
\newcommand{\gbnd}{g_{0}^{b}}
\newcommand{\hbnd}{h_{0}^{b}}
\newcommand{\zh}{z_{h}}
\newcommand{\zb}{z_{b}}
\newcommand{\man}{\mathcal{M}}
\newcommand{\hbr}{\bar{h}}
\newcommand{\tbr}{\bar{t}}

\begin{titlepage}
\begin{flushright}
CHEP XXXXX
\end{flushright}
\bigskip
\def\thefootnote{\fnsymbol{footnote}}

\begin{center}
{\Large
{\bf Hairy Black Holes in a Box \\ \vspace{0.1in} 
}
}
\end{center}

\bigskip
\begin{center}
{\large Pallab BASU$^a$\footnote{\texttt{pallabbasu@gmail.com}},  Chethan KRISHNAN$^b$\footnote{\texttt{chethan.krishnan@gmail.com}} and  P. N. Bala SUBRAMANIAN$^b$\footnote{\texttt{pnbala@cts.iisc.ernet.in}}}
\vspace{0.1in}

\end{center}

\renewcommand{\thefootnote}{\arabic{footnote}}

\begin{center}
$^a$ {International Center for Theoretical Sciences,\\
IISc Campus, Bangalore 560012, India}\\
\vspace{0.2in}

$^b$ {Center for High Energy Physics,\\
Indian Institute of Science, Bangalore 560012, India}\\

\end{center}

\noindent
\begin{center} {\bf Abstract} \end{center}

We do a systematic study of the phases of gravity coupled to an electromagnetic field and charged scalar in flat space, with box boundary conditions. The scalar-less box has previously been investigated by Braden, Brown, Whiting and York (and others) before AdS/CFT and we elaborate and extend their results in a language more familiar from holography.   The phase diagram of the system is analogous to that of AdS black holes, but we emphasize the differences and explain their origin. Once the scalar is added, we show that the system admits both boson stars as well as hairy black holes as solutions, providing yet another way to evade flat space no-hair theorems. Furthermore both these solutions can exist as stable phases in regions of the phase diagram. The final picture of the phases that emerges is strikingly similar to that found recently for holographic superconductors in global AdS, arXiv: 1602.07211. Our construction lays bare certain previously unnoticed subtleties associated to the definition quasi-local charges for gravitating scalar fields in finite regions. 

\vspace{1.6 cm}
\vfill

\end{titlepage}

\setcounter{footnote}{0}

\section{Introduction}

Schwarzschild black holes in flat space have negative specific heat, which means that they heat up by Hawking radiating and cool down by absorbing radiation. Therefore, they cannot be in equilibrium with thermal radiation in asymptotically flat space. As is well known, one way to bypass this problem is to put the black hole in a (small enough) box, and to study the phases of the black hole + radiation system. 

A natural gravitational box for the black hole is provided by Anti-de Sitter space, where the phase structure of pure gravity was studied for the the first time by Hawking and Page \cite{HP}. With the advent of the AdS/CFT correspondence \cite{Malda, GKP, Witten1, Witten2}, it became clear that this is more than just a curiosity and that the physics of the black hole in the AdS box is dual to that of a (de-)confined gauge theory. 

AdS/CFT correspondence triggered an avalanche of interest, and black holes in (asymptotically) AdS geometries have been studied from various angles. In particular it has been noted that adding a charged scalar to the Einstein-Maxwell system in AdS gives a way to evade the no-hair theorems of flat space\footnote{Black holes are essentially uniquely determined by their global charges: this is the basic message of the no-hair theorems in classical general relativity. The {\em spirit} of these theorems is unlikely to be evaded, and forms one of the cornerstones of the modern lore on black holes: we need a quantum count of the microstates of black holes to account for the Bekenstein-Hawking entropy, classical hair simply is not numerous enough. But the {\em letter} of the no-hair theorems have been evaded in many ways, and holographic superconductors in AdS are an example. Our box black holes will serve as another.}: in the AdS/CFT literature such black holes are called holographic superconductors \cite{HHH1, HHH2} for reasons that we will not delve into. The detailed study of the phase structure and other properties of this system and its numerous generalizations have given rise to an industry in itself \cite{BKA,BasuVector,IIB,HorIntro}. 

However, the original question that motivated Hawking and Page to consider AdS space in the first place, namely the black hole in a box, has {\em not} been investigated much in the context of the added luxury of a charged scalar. In particular, the phase diagram of the Einstein-Maxwell-scalar system in a box is not known, to the best of our knowledge. Our goal in this paper is to take a first step in that direction and to chart out the phase diagram of this system. We will find that apart from the known Reissner-Nordstrom black hole, the system also allows boson stars and hairy black holes as classical solutions. We will furthermore demonstrate that these solutions are more than a curiosity: they exist as thermodynamically stable phases in appropriate regions of the $T-\mu$ phase diagram ($T$ is the temperature and $\mu$ is the chemical potential of the system). Our construction of hairy solutions is a constructive proof for yet another way to evade the no-hair theorems of asymptotically flat space. 

The phase diagram that we uncover bears striking resemblance to that of the Einstein-Maxwell-scalar system in global AdS studied in \cite{BKS1602} (see also \cite{Arias}). This is re-assuring because our expectation is that AdS should really be viewed as a box. We work with the specific case of the massless scalar for concreteness, and our comparison will be with the conformally coupled scalar studied in \cite{BKS1602}.

In what follows, we first start out by considering the Einstein-Maxwell system (without a scalar) in a box. This system (as well as the pure Einstein system \cite{PHut,GibbonsPerry}) have been studied before and our results will overlap with those of Braden, Brown, Whiting and York \cite{BBWY}. Our approach will however be decidedly holography-inspired and somewhat more complete. We will also emphasize the definitions of the charges etc., which will have to be reconsidered when we add the charged scalar. Adding the scalar brings in a few different subtleties, related to the fact that no-hair theorems are in effect when the box size is taken to infinity. This also introduces difficulty in defining quasi-local charges directly, as we will discuss. But the free energy is well-defined and computable and gives rise to a phase diagram that matches with our qualitative expectations from global AdS \cite{BKS1602} as well as reduces to the hairless case when the scalar is turned off. We will conclude with some comments and possible future directions for work.

{\bf Note Added:} After this paper was substantially completed, we
became aware of some results in the literature where hairy solutions
in a cavity have been constructed before, most notably \cite{Win1, Win2, Win3}.
Isolated special examples of hairy solutions were shown to
arise even earlier as the final states of super-radiant instabilities
in \cite{Herdiero}, see also \cite{Bosch}\footnote{Some of the papers in \cite{Herdiero} were looking at the growth of the scalar field in the linear regime only, where it grows exponentially. So these were not true solutions to the field equations, but indicative.}. A relevant conjecture here
is that of \cite{HerdeiroRadu}. These observations indicate that these
solutions can arise as the endpoints of dynamical processes,
suggesting that they can be stable and physical. This is
satisfying, in light of the results of our work: we deal with the stability of thermal phases, these papers deal
with dynamical aspects.

The work of \cite{Win1, Win2, Win3} offers a nice complementary
discussion\footnote{The solutions they find seem identical to ours modulo
notations and conventions, except for one caveat: we have had to be
somewhat more careful with boundary issues than \cite{Win1, Win2,
Win3} for various reasons. To make the box boundary fully well-defined as
a variational problem, one needs to add
a boundary term to the action (the Gibbons-Hawking-York term, see our
discussion in Section 3). This means that the problem is well-defined
only with a fixed boundary metric, which we take as our ``box"
(\ref{bndmet}), and we write down all our bulk solutions in the {\em same}
gauge for the boundary metric, namely (\ref{bndmet}). Instead, \cite{Win1,Win2, Win3} hold $h(r)$ fixed to unity at the {\em horizon}, which means that they will need a further {\em solution-dependent} rescaling of the time coordinate that brings the boundary value of $g(r) h(r)$ to some fixed value (say unity) to bring all their solutions into the same gauge.  The $ g(r) $ and $ h(r) $ here are defined in eg. \eqref{metric}.} of these solutions: our focus is on thermodynamic stability, they focus on perturbative stability. Taking these results together, it seems evident that these solutions are bonafide solutions of gravity in a box.
\section{The Setup}

We will consider a spacetime manifold $\man$, with a time-like boundary $\partial\man$, which we will refer to as a {\em box} henceforth. We will first look at gravity, with no cosmological constant, coupled to Maxwell field: the intuition we get by studying this system will be useful when we add the scalar in later sections. The action is given by
\bea\label{action}
S=  \frac{1}{16 \pi}\int_{\man} d^{4}x \sqrt{-g}\left( R - F_{\mu\nu} F^{\mu\nu}\right) + \dfrac{1}{8\pi }\oint_{\partial \mathcal{M}} \sqrt{-\gamma} \; \mathcal{K} \, ,
\eea
where $g_{\mu\nu}$ gives the metric in the bulk, $\gamma$ is the metric in the boundary, and $\mathcal{K}$ is the extrinsic curvature. The boundary piece in the action is called the Gibbons-Hawking-York term, and we will briefly comment about it in the next section. We have set $G=1$. Note that the normalization of our Maxwell piece follows the conventions of \cite{catastrophic}.

We would like to work with a time independent ansatz, which is also spherically symmetric.  We will be looking at a space where the boundary is at $ r = \rb $. The metric is chosen to be of the form
\bea \label{metric}
ds^{2} = -g(r) h(r) dt^{2} + \dfrac{dr^{2}}{g(r)} + r^{2}\, d\Omega_{2}^{2},
\eea
and for the Maxwell field (see similar constructions in eg. \cite{HHH2, BKA})
\bea
A = \phi(r)dt.
\eea
With the above ansatz, we get the equations of motion
\begin{align}
\label{eoms}
&\frac{g'(r)}{r g(r)}+\frac{\phi '(r)^2}{g(r) h(r)}-\frac{1}{r^2 g(r)}+\frac{1}{r^2} =0, \\
&h'(r) = 0, \\
&\phi ''(r)+\frac{2 \phi '(r)}{r} -\frac{h'(r) \phi '(r)}{2 h(r)} = 0.
\end{align}
The second of the above equations is solved by $h(r)$ a constant, but we will phrase the discussion below at the level of the equations of motion without setting $h(r)$ to constant. The reason for doing this is that when one adds the scalar, the $h$-equation of motion will become non-trivial (see later sections), but the discussion below will still hold. 
From the equations of motion, we can see the existence of the following two scaling symmetries.
\begin{itemize}
\item $r \rightarrow a r$. With this rescaling, one could set $\rb  =1$.
\item $h \rightarrow \bar{h}= a^{2} h, \ \phi \rightarrow \bar{\phi} =  a \phi, \ {\rm and }\ t \rightarrow \bar{t}= \frac{t}{a}$. This scaling symmetry can be used to set the $g_{tt}$ coefficient of the metric to be unity at $r=\rb $. The boundary metric will thus be $\mathbf{R} \times \mathbf{S}^{2}$, which will ensure that the metric of any geometry matches with the flat space metric at the boundary. This gives
\bea
-g h dt^{2}|_{\rb} = - g \dfrac{\bar{h}}{a^{2}}dt^{2}|_{\rb} = -g \bar{h} d\bar{t}^{2}|_{\rb} = -d\bar{t}^{2}.
\eea
For this, we choose rescaling in the following way\footnote{In the rest of the paper, we will drop the {\em bar} on the variables $\bar{h}$ and $\bar{t}$, with the understanding that the boundary conditions are met}.
\bea
\lim_{r\rightarrow\rb }\bar{h}(r) = \dfrac{1}{g(\rb)}, \ \text{i.e} \ \ \bar{h}(\rb)= a^{2} h(\rb)=\dfrac{1}{g(\rb)} \Rightarrow a = \dfrac{1}{\sqrt{g(\rb )h(\rb )}},
\eea
\end{itemize} 
We will be interested in looking at the Schwarzschild and Reissner-Nordstr\"om solutions, inside the box. 

\section{Gravity in a Box}

Because gravity is that mysterious force that causes spacetime itself to be dynamical, putting gravity in a box strikes terror in the hearts of some. What if gravitational waves leak out uncontrollably beyond the box boundary? Occasionally, such worries were voiced when the authors gave talks on related topics before this paper was completed. So let us start by making a few comments to assuage such fears. First of all, a dynamical spacetime does {\em not} mean that there is  any violence done to the manifold structure: it means merely that metric is the dynamical variable. Secondly, by putting gravity in a box, what one operationally does, is to set an appropriate boundary condition for the metric. And despite the fact that it affects our notions of distance and is therefore sacred to us, on a manifold the metric is just {\em some} field. This means that at least classically, the ``box boundary condition'' is perfectly well-defined as a boundary condition for the metric, as long as the metric equations of motion can arise from a well-posed variational problem on the manifold, with the said ``box boundary condition''. In particular, gravitational waves cannot do anything illegal if this is the case, because gravitational waves are solutions of the metric equations of motion that arise from such a variational problem, and therefore {\em by construction} have to respect those boundary conditions.

So in order for the ``box boundary condition'' to be physically acceptable, what we need to make sure is that they lead to a well-defined variational formulation for the metric. Now the most natural ``box boundary condition" is to hold the metric at the boundary fixed\footnote{A ``box" is nothing but a Dirichlet boundary condition for the fields.}, but it is known ever since the work of Gibbons-Hawking \cite{GH} and York \cite{York} that Einstein gravity indeed allows a perfectly well-defined variational problem of this Dirichlet type, when one adds the so-called Gibbons-Hawking-York boundary term to it. So this is the reason why we work with an action of the form (\ref{action}) in this work (and its generalization to include a charged scalar which we will consider a bit later). We also note that the standard scalar and Maxwell pieces in the Lagrangian automatically are well-defined Dirichlet problems, so we do not need to add any boundary terms for them.

In what follows, for all geometries with or without a horizon, we will take the boundary metric to be of the form 
\bea\label{bndmet}
ds^{2}|_{\partial\man} = -dt^{2} + \rb^{2} d\Omega_{2}^{2}.
\eea
This is our definition of the box. This means in particular that $g_{tt}^{black hole}|_{\rb } = g_{tt}^{flat space}|_{\rb } $, so  the effective temperatures defined for the two systems will be equal at $r= \rb$. This will be relevant if/when we do background subtraction of the classical action of non-trivial geometries with that of flat space.  As the manifold does not have an asymptotic region, the definitions of the energy, charge and temperature require a bit of explanation for those who are used to asymptotically flat/AdS boundary conditions. 

The Hawking temperature computation is unaffected because it relies only on the horizon and not the boundary. When there is a horizon at $r=\rh$, the line element (\ref{metric}) can be expanded via $r = \rh + \delta r$ and after the usual \cite{QFTBH} demand that there are no conical singularities in the Euclidean metric, one ends up with  
\bea\label{tbh}
\dfrac{1}{\beta} = T = \dfrac{1}{4\pi} g'(\rh)\,h(\rh)^{1/2}
\eea
as the Hawking temperature. The entropy of the geometry is also a horizon quantity, and is just given by the a quarter of the area of the horizon as usual:
\bea
S=\pi \rh^2.
\eea

In order to define the gravitational mass in an asymptotically flat spacetime, we can use the ADM construction. For defining the ADM mass, the spacelike slices ($\Sigma_{t}$) of the geometry are set up in such a way that they asymptotically coincide with a constant time surface of Minkowski space. The spacelike slices $\Sigma_{t}$ are bounded by closed two-surfaces $S_{t}$.  The mass is then defined as the value of the ADM Hamiltonian when the two-surface is a two-sphere at spatial infinity, for a specific choice of lapse and shift\footnote{See Sec.4.3 of \cite{Poisson} for details, the ADM mass is obtained when the lapse is taken to be unity and the shift is taken as zero. This identifies the ADM mass as the generator of boundary time translations.}, as
\bea
M = -\dfrac{1}{8\pi} \lim_{S_{t}\rightarrow\infty} \oint_{S_{t}} (k - k_{0})\sqrt{\sigma}d^{2}\theta,
\eea
where $\sigma_{AB}$ is the metric on $S_{t}$, $k$ is the extrinsic curvature of $S_{t}$ embedded in $\Sigma_{t}$ and $k_{0}$ is the extrinsic curvature of $S_{t}$ embedded in flat space. Using this definition, we find that the ADM mass of a black hole is the mass parameter $M$ of the Schwarzschild metric, and it has the interpretation of energy in the thermodynamics of the system, i.e. $E = M $.

In our construction, the space does not have an asymptotic region, instead, we set the spacelike slices $\bar{\Sigma}_{t}$ to be in such a way that the boundary metric coincides with a constant time slice of Minkowski metric with a boundary at $r=\rb$. The quasilocal energy density can be defined as \cite{BrownYork}
\bea\label{energy}
E \equiv -\dfrac{1}{8\pi} \lim_{r\rightarrow\rb } \oint_{\mathbf{S}^{2}} (k - k_{0})\; \sqrt{\sigma} \; d^{2}\theta.
\eea
where $\sigma_{AB} = \rb^{2} d\Omega^{2}_{2}$ is the metric of the boundary 2-sphere, and the unit normal is $n_{\mu} = (0,g(r_{b})^{-1/2},0,0)$. The extrinsic curvature of the boundary $2$-sphere, of the geometry embedded in the spacelike slices $\bar{\Sigma}_{t}$ is given by $k$, and $k_{0}$ is the extrinsic of the boundary $2$-sphere embedded in flat space. Also, $k = k_{AB}\sigma^{AB}$, and 
\bea
k_{AB} = \dfrac{1}{2} (\nabla_{\mu} n_{\nu}+\nabla_{\nu} n_{\mu}) e^{\mu}_{A} e^{\nu}_{B},
\eea
where $e^{\mu}_{A}= \partial x^{\mu}/ \partial \theta^{A}$ denote the basis vectors of the $2$-sphere.

The definition of chemical potential is taken to be the value of the gauge field potential at the boundary, $\mu = \phi(\rb) $. The electric charge of the system is defined as 
\bea
Q = \lim_{r\rightarrow \rb} \dfrac{1}{4\pi}\int_{S^{2}} F_{\mu\nu} t^{\mu}n^{\nu} \sqrt{\sigma} d^{2}\theta,
\eea
where $t^{\mu}$ is the unit time-like normal at the boundary, and $n^{\nu}$ is the unit outward drawn normal at the $r=\rb $ hypersurface. 

To evaluate the classical action for the geometries directly, we have to know the boundary term or the Gibbons-Hawking term. For the metric ansatz we have chosen, the outward unit normal to $\partial\man$ is given by $n_{\mu} = ( 0 ,g(\rb)^{-1/2},0,0)$. The metric at the boundary, after appropriate rescaling, is $\gamma_{IJ}\,dy^{I}\,dy^{J} = -dt^{2} +\rb^{2}\,d\Omega_{2}^{2} $, where $y^{I} = (t,\theta,\phi)$. The extrinsic curvature of the boundary embedded in the full geometry is given by $\mathcal{K} = \mathcal{K}_{IJ}\gamma^{IJ}$, where
\bea
\mathcal{K}_{IJ} = \dfrac{1}{2}(\nabla_{\mu}\eta_{\nu} + \nabla_{\nu}\eta_{\mu}) e^{\mu}_{I}e^{\nu}_{J}
\eea
and $e^{\mu}_{I}= \partial x^{\mu}/ \partial y^{I}$ are the basis vectors at $\partial\man$. Evaluating this for our metric ansatz, we get
\bea
\mathcal{K} = \frac{1}{2 \sqrt{g(\rb)}}  g'(\rb)+ \dfrac{\sqrt{g(\rb)} h'(\rb)}{2\, h(\rb)}+\frac{2 \sqrt{g(\rb)}}{\rb}.
\eea
The extrinsic curvature for the Minkowski box will be denoted as $\mathcal{K}_{0}$, and will be used to do a background subtraction, which sets the free energy of the Minkowski box to zero. The background subtraction is strictly not necessary if we are looking at the box, as there are no divergences. However, doing a background subtraction makes the comparison of quantities more straightforward when we want to take the limit when the boundary goes to infinity and we hope to reproduce the known results in asymptotically flat space\footnote{When there is a non-trivial scalar profile in the problem, that there is no such smooth asymptotically flat limit, is one of the observations of this paper.}. 

\section{Schwarzschild in the Box}

The simplest non-trivial solution for the equations of motion in \eqref{eoms} is given by,
\bea
h(r) = C_{1},\ \ \phi(r) = \mu, \ \text{and}\ g(r) = 1- \dfrac{\rh}{r},
\eea
where $C_{1}$ is some constant, which we will set to be $1/g(\rb)$, and $\mu$ is a constant chemical potential, which is arbitrary for the Schwarzschild solution. We will set it to zero for convenience, because it does not affect the following discussion.

\begin{figure}[H]
\centering
\includegraphics[width=0.5\textwidth]{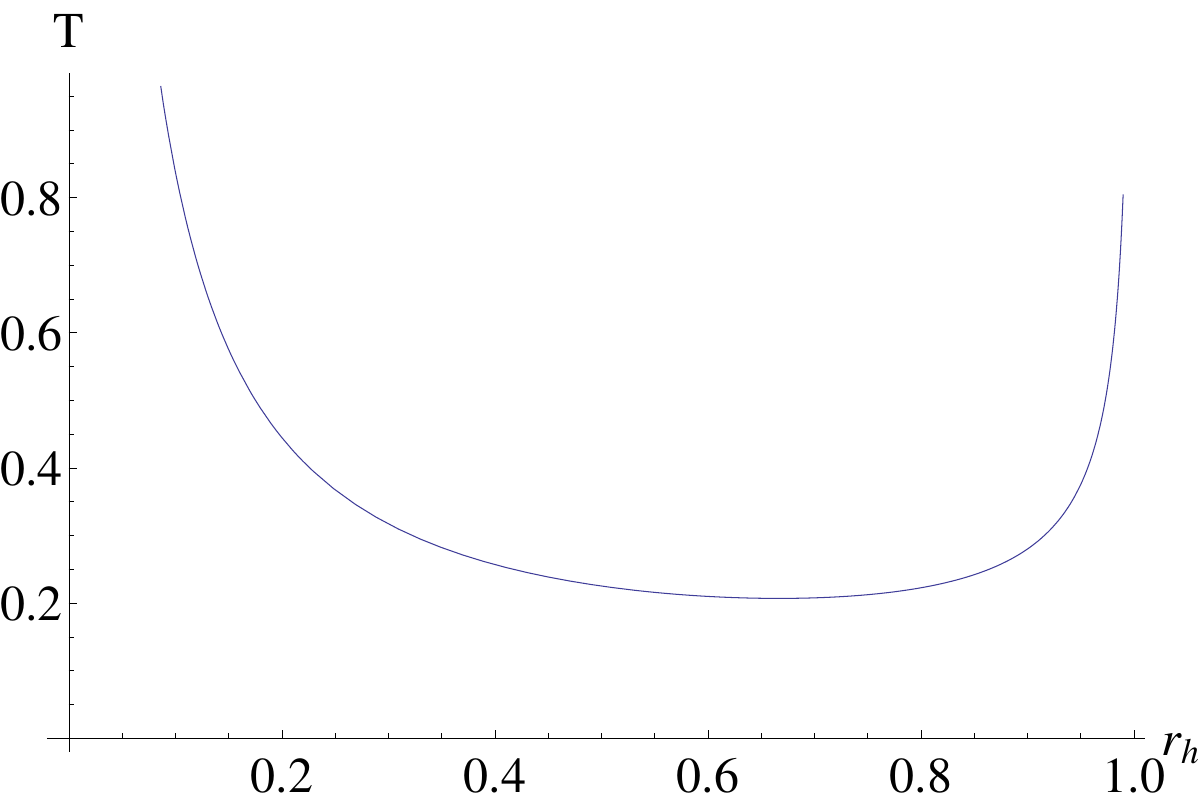}
\caption{Temperature as a function of $\rh $, with $\rb =1 $.}
\label{tempsch}
\end{figure}

The quasilocal energy and the temperature of the Schwarzschild solution can be computed as described above:
\bea
&E = \rb -\rb  \sqrt{1-\dfrac{\rh}{\rb}},\\
&T =\dfrac{1}{4\pi } \dfrac{1}{\rh \sqrt{g(\rb)}} = \dfrac{1}{4\pi \,\rh }\sqrt{\dfrac{\rb }{\rb -\rh }}
\eea
The temperature is plotted as a function of $\rh $, after setting $\rb = 1$, in Fig.\ref{tempsch}. It can be seen from the figure that the temp for a very small black hole and a black hole approaching the size of the box go off to infinity, and for any temperature above $T_{min}$ there are two black hole solutions.
The free energy of the system is given by 
\bea
F &= E - T S = \rb -\rb  \sqrt{1-\dfrac{\rh}{\rb}} -\dfrac{1}{4\pi}\dfrac{\sqrt{\rb} }{\rh \sqrt{\rb - \rh }} \pi \rh ^{2} \nonumber\\
&= \rb -\rb  \sqrt{1-\dfrac{\rh}{\rb}} -\dfrac{\sqrt{\rb} \rh }{4 \sqrt{\rb - \rh }} .
\eea

The free energy can also be computed directly from the classical action. The only term that will contribute is the surface term, because the Ricci scalar $R=0$ and the gauge field is not turned on.
\bea
F = T S_{cl} = - \dfrac{T}{8\pi} \int_{0}^{1/T}d\tau \int d\theta \, d\phi \, \sin^{2}\theta \, r^{2} \left( \mathcal{K} - \mathcal{K}_{0}\right)\biggr|_{\rb },
\eea
The free energy computed using this formula yields the same result, and we also have $\frac{\partial E}{\partial S} = T$.

\begin{figure}[H]
\centering
\includegraphics[width=0.5\textwidth]{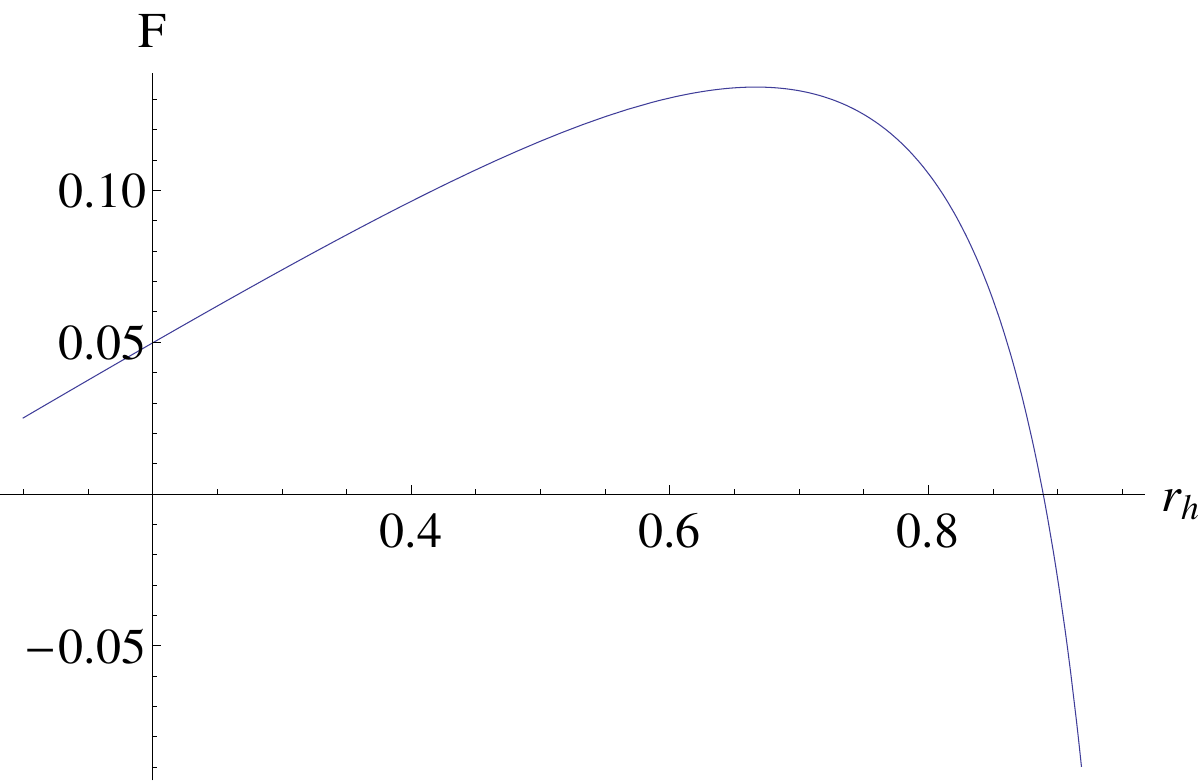}
\caption{Free energy as a function of $\rh $, with $\rb =1 $.}
\label{schferg}
\end{figure}

\begin{figure}[H]
\centering
\includegraphics[width=0.5\textwidth]{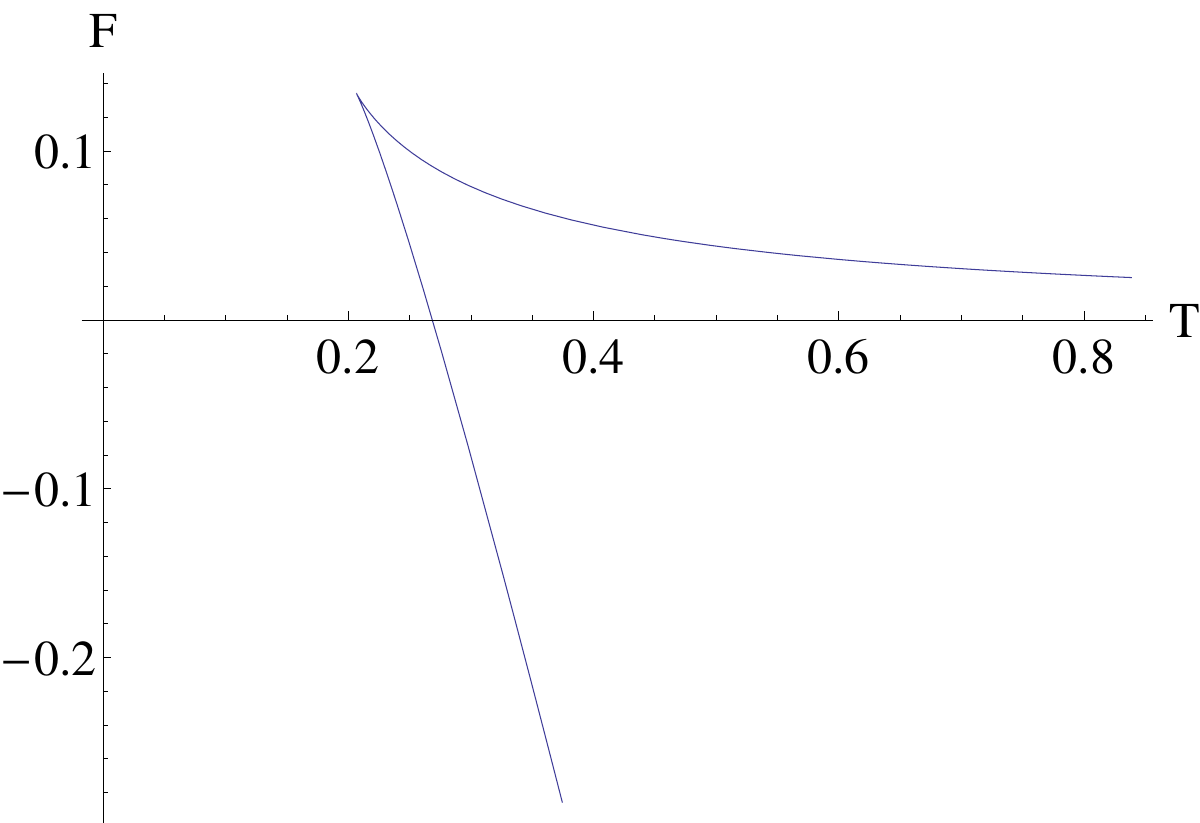}
\caption{Free energy plotted against $T$, with $\rb =1 $.}
\label{schfergt}
\end{figure}

In Fig.\ref{schferg}, we have plotted the free energy of the system against $\rh $, and against $T$ in Fig.\ref{schfergt}, with $\rb =1$. The free energy is positive for a small black hole, and goes negative for black hole larger than $\rh =\frac{8}{9}\rb $, which is the box analogue of the AdS Hawking-Page transition.

The plots in Fig.\ref{schferg},\ref{schfergt} looks similar to the Schwarzschild black hole in global AdS, and so do the Penrose diagrams of Schwarzschild black hole in global AdS and in a box (with no cosmological constant)\footnote{See http://www.iopb.res.in/$\sim$mukherji/THESIS/tanay.pdf figures 2.1, 2.2, 2.3 for the Schwarzschild-AdS black hole}. 

\section{Reissner-Nordstr\"om in the Box}

Now, we will look at the RN solution in the box. We solve for $g(r),h(r),\phi(r)$ for the equations of motion in \eqref{eoms} and appropriately rescale them to get
\bea
\begin{split}
g(r) &= 1-\dfrac{Q^2+\rh ^2}{\rh \, r }+\dfrac{Q^2}{r^2} = 1 - \dfrac{(1 + \epsilon) \rh }{r} + \dfrac{\epsilon \rh^{2}}{r^{2} }, \\ \hbr(r) &= \dfrac{1}{g(\rb)}, \ \text{and} \ \bar{\phi}(r) = \dfrac{Q}{\sqrt{g(\rb)}}\left(\dfrac{1}{\rh }- \dfrac{1}{r}\right) = \dfrac{\sqrt{\epsilon}\, \rh }{\sqrt{g(\rb)}}\left(\dfrac{1}{\rh }- \dfrac{1}{r}\right),
\end{split}
\eea
where we have parametrized the inner horizon as 
\bea\label{rinner}
r_{inner} = \frac{Q^{2}}{\rh} = \epsilon \rh 
\eea
with $0\leqslant \epsilon \leqslant 1$.
The energy and temperature of the system can again be computed:
\bea
E = \rb - \rb  \sqrt{1 -\frac{(1 + \epsilon )\rh  }{\rb } + \frac{\epsilon  \,\rh^{2}}{\rb ^2}} ,\\
T= \dfrac{1}{4\pi} \dfrac{(1 - \epsilon^{2})}{\rh }\left(1 -\frac{(1 + \epsilon )\rh  }{\rb } + \frac{\epsilon \, \rh^{2}}{\rb ^2}\right)^{-1/2}
\eea
The chemical potential of the system is 
\bea
\mu = \bar{\phi}(\rb) = \dfrac{\sqrt{\epsilon}\, \rh}{\sqrt{g(\rb)}} \left(\dfrac{1}{\rh }- \dfrac{1}{\rb }\right).
\eea

The thermodynamic relations $T= \frac{\partial E}{\partial S }\bigr|_{Q }$ and $\mu = \frac{\partial E}{\partial Q }\bigr|_{\rh }$ can be checked to hold from these. Putting all this together we get the free energy
\bea
F = E - T\, S - \mu \, Q \hspace{10.5cm}\nonumber \\
= \left(\rb\sqrt{1 -\frac{(1 + \epsilon )\rh  }{\rb } + \frac{\epsilon \rh^{2}}{\rb ^2}} - \rb +\dfrac{\epsilon  \,\rh }{4 } +\dfrac{3\rh }{4} \right) \left(1 -\frac{(1 + \epsilon )\rh  }{\rb } + \frac{\epsilon \rh^{2}}{\rb ^2}\right)^{-1/2}
\eea
The expressions for $E, T$ and $\mu$ that we obtain are the same as that in \cite{BBWY}, and also of the on-shell action. In \cite{BBWY}, the analysis is centered around finding configurations that are locally stable, although, they point out that certain configurations give a global minima for the on-shell action. We will systematically analyze the phase structure of the RN black hole in a box, using the free energy to characterize the thermodynamic stability, which is the language that is familiar from AdS-CFT.

For $\epsilon=0$, we must get the Schwarzschild case, and the black hole will be extremal when $\epsilon =1$. The free energy set to zero gives the transition curve between flat space and the RN black hole. This can be computed fully analytically, and we get the solutions
\bea
\epsilon = 1 , \; \dfrac{9 \rh - 8\rb }{\rh }.
\eea

For the (not-so-interesting) case with $\epsilon =1$, the black hole is extremal and will remain so for any value of $\rh < \rb$. The chemical potential and temperature are given by 
\bea
\mu =1  \ \text{and} \ T = 0.
\eea

\begin{figure}[H]
\centering
\includegraphics[width=0.5\textwidth]{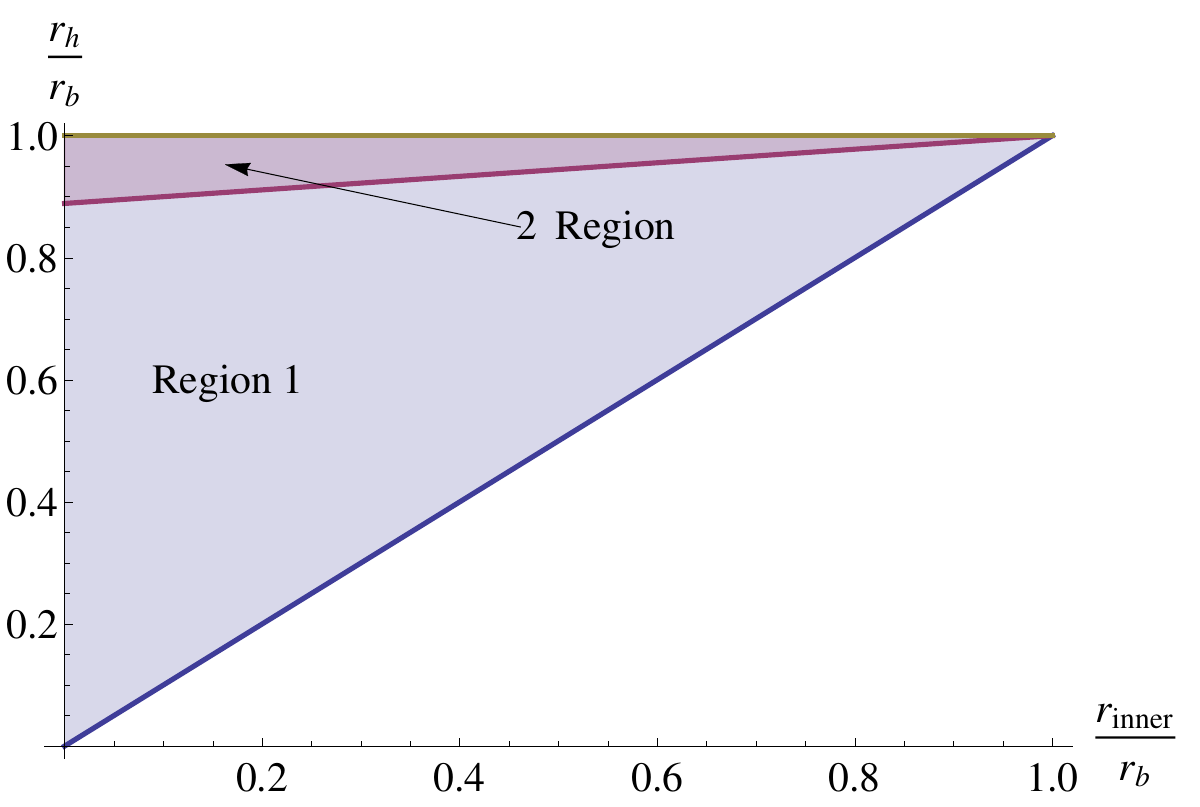}
\caption{Region 1 and Region 2 together indicate where black holes can be formed and Region 2 is where they are thermodynamically favourable. See \eqref{rinner} for the definition of $r_{inner}$.}
\label{rnbhregion}
\end{figure}

For the case $\epsilon = \frac{9 \rh - 8\rb }{\rh }$, the chemical potential and temperature are given by
\bea
\mu =\sqrt{1-\frac{8 r_b}{9 r_h}}\ \ \text{and} \  T= \dfrac{2\, \rb }{3\pi \rh^{2}}.
\eea
In the case of global AdS, as we look at larger values of $\mu$ along the Hawking-Page transition curve, the horizon continues to shrink, and intersects the $T=0$ axis at $\mu=1$ (see eg., \cite{BKS1602}). However, in the case of the RN black hole in the box, the black hole gets bigger and bigger as we go up in $\mu$, and gets closer to extremality as the black hole becomes almost the size of the box itself, as can be seen from the following relation
\bea
\dfrac{\rh }{\rb } = \dfrac{8 }{9(1-\mu^{2})} \leqslant 1.
\eea
In Fig.\ref{rnbhregion}, we have shown the the regions where the RN black hole can exist in the box, and also on where it becomes thermodynamically favorable. 

\begin{figure}[H]
\centering
\includegraphics[width=0.5\textwidth]{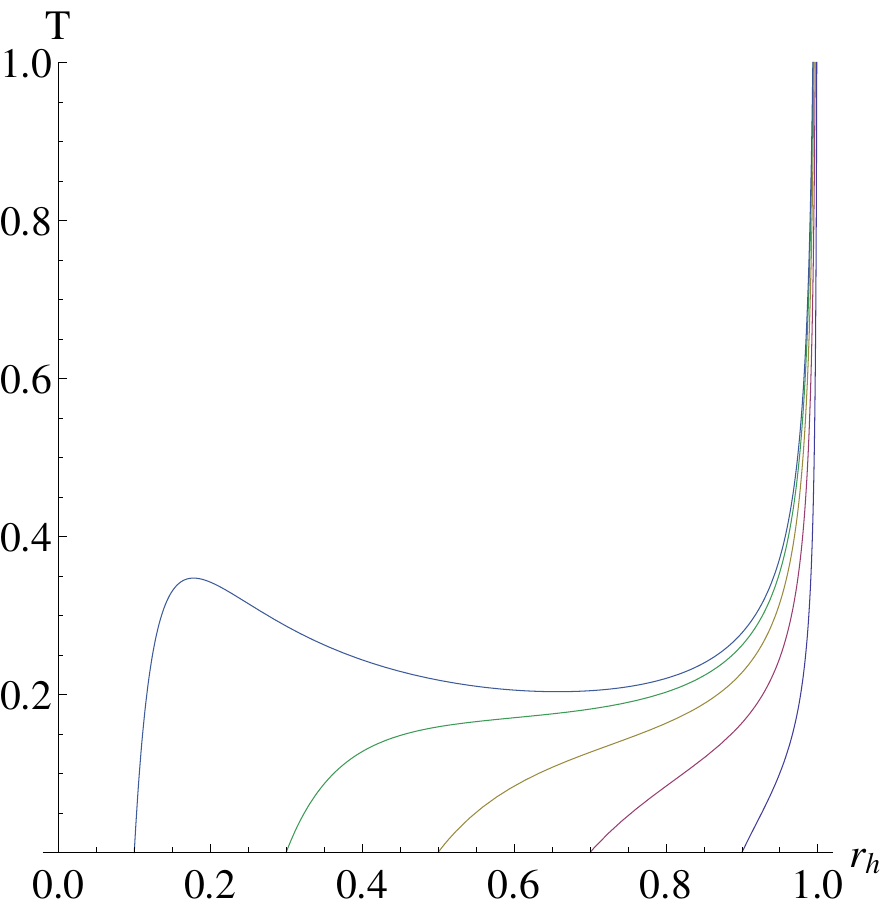}
\caption{Temperature of RNBH with different $Q$ against $\rh $, with $\rb =1 $, for $Q=0.1,0.3,0.5,0.7,0.9$ (in that order from left to right).}
\label{compare}
\end{figure}

At $\rh = \frac{8}{9}\rb $, we will get $\epsilon = 0$ in the second case, which corresponds to $\mu =0$, and $T = \frac{27}{32\pi \rb}$, which is the Schwarzschild case. The more interesting limit happens at $\rh \rightarrow \rb$. The chemical potential becomes $\mu=\frac{1}{3}$, and temperature $T = \frac{2}{3\pi}$ along with $\epsilon \rightarrow 1$. This means the phase diagram will have an abrupt ending at some finite $T$. The reason this happens can be understood as follows. As the outer horizon of the black hole is very near the boundary, the temperature diverges, see Fig.\ref{compare}. However, along with that, to make the free energy zero, the inner horizon is approaching the outer horizon, making it almost extremal, and it tries to take the temperature to almost zero. The existence of a finite limit is a balancing of this competition.

\begin{figure}[H]
\centering
\includegraphics[width=0.6\textwidth]{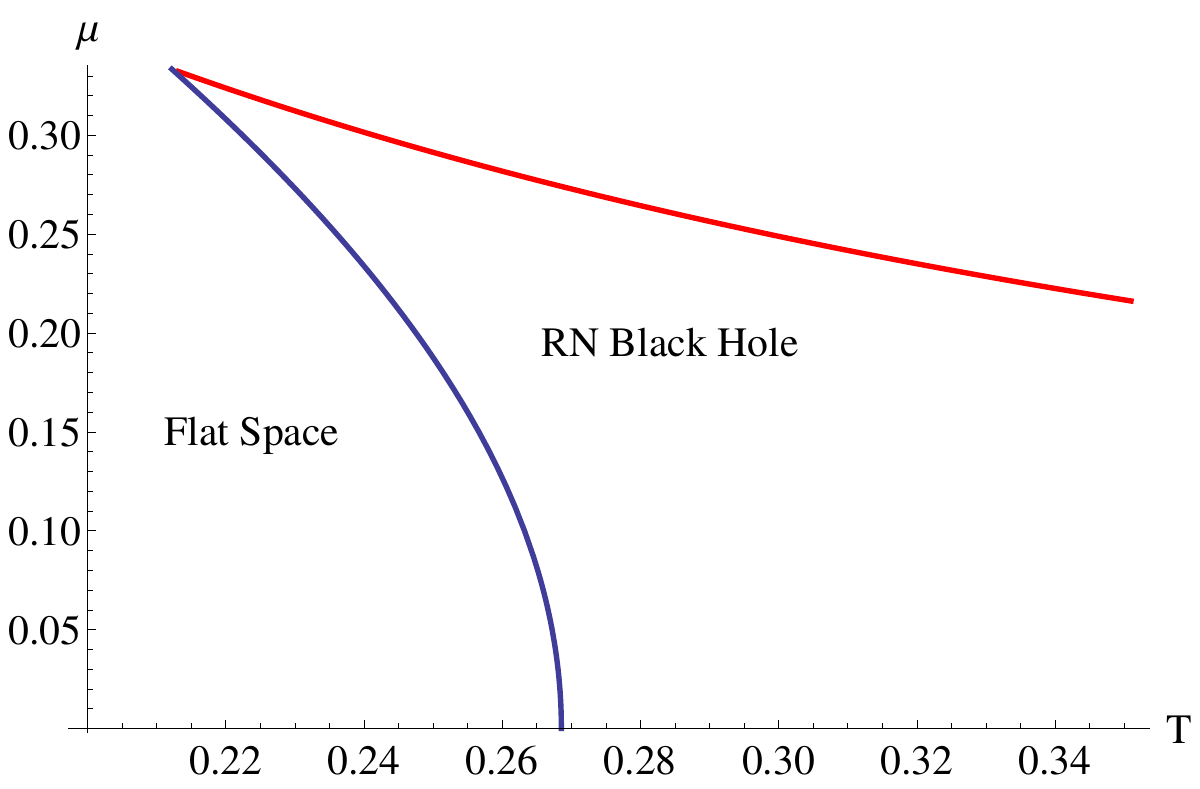}
\caption{Phase diagram of RN BH in a box.}
\label{rnphasedig}
\end{figure}

In Fig.\ref{rnphasedig}, this curve is shown in blue. This curve, as it can be seen has an abrupt ending, at $T=\frac{2}{3\pi} = 0.2122$. However, the big RN black hole phase has another phase boundary, which comes from the saturation of the box itself, i.e. the black hole horizon approaching the size of the full box. At the $\mu = 0$ axis, this will be at $T \rightarrow \infty$, and it is a Schwarzschild black hole limit.

To understand the behaviour of a box-sized near-extremal black hole, let us look at the expressions for $T$ and $\mu$ in this limit. First, we will parametrize $\epsilon = 1 - \delta $, where $\delta \ll 1$. Now, in this limit we get
\bea
T = \dfrac{1}{4\pi \rh } \left(\left[\dfrac{\rh \delta}{1-\rh}\right] + \dfrac{1}{2}\left[\dfrac{\rh \delta}{1-\rh}\right]^{2} + \dfrac{3}{8}\left[\dfrac{\rh \delta}{1-\rh}\right]^{3} +\dots\right),\\
\mu = 1 - \dfrac{1}{2} \left(\dfrac{\delta}{1-\rh }\right) -\dfrac{1-4\rh }{8}\left(\dfrac{\delta}{1-\rh }\right)^{2} -\dfrac{1-4\rh +8\rh^{2}}{16} \left(\dfrac{\delta}{1-\rh }\right)^{3} + \dots.
\eea
From here one can see that if $\rh$ not close to $1$, then in the $\delta\rightarrow 0$ limit, or $\epsilon \rightarrow 1$, the temperature will go to $0$, and $\mu = 1$. However, if $\rh\rightarrow 1$ at the same time then what we will end up is a limit of the form
\bea
\lim_{\delta\rightarrow 0 }\lim_{\rh\rightarrow 1} \dfrac{\delta}{1-\rh}. \nonumber
\eea
This appears in both $T$ and $\mu$ expansions and is what gives the finite temperature limit for the almost box-sized near-extremal black holes. The red curve in Fig.\ref{rnphasedig} is the plot for a almost-extremal black hole that is infinitesimally smaller than the box itself.

\begin{figure}[H]
\centering
\includegraphics[width=0.5\textwidth]{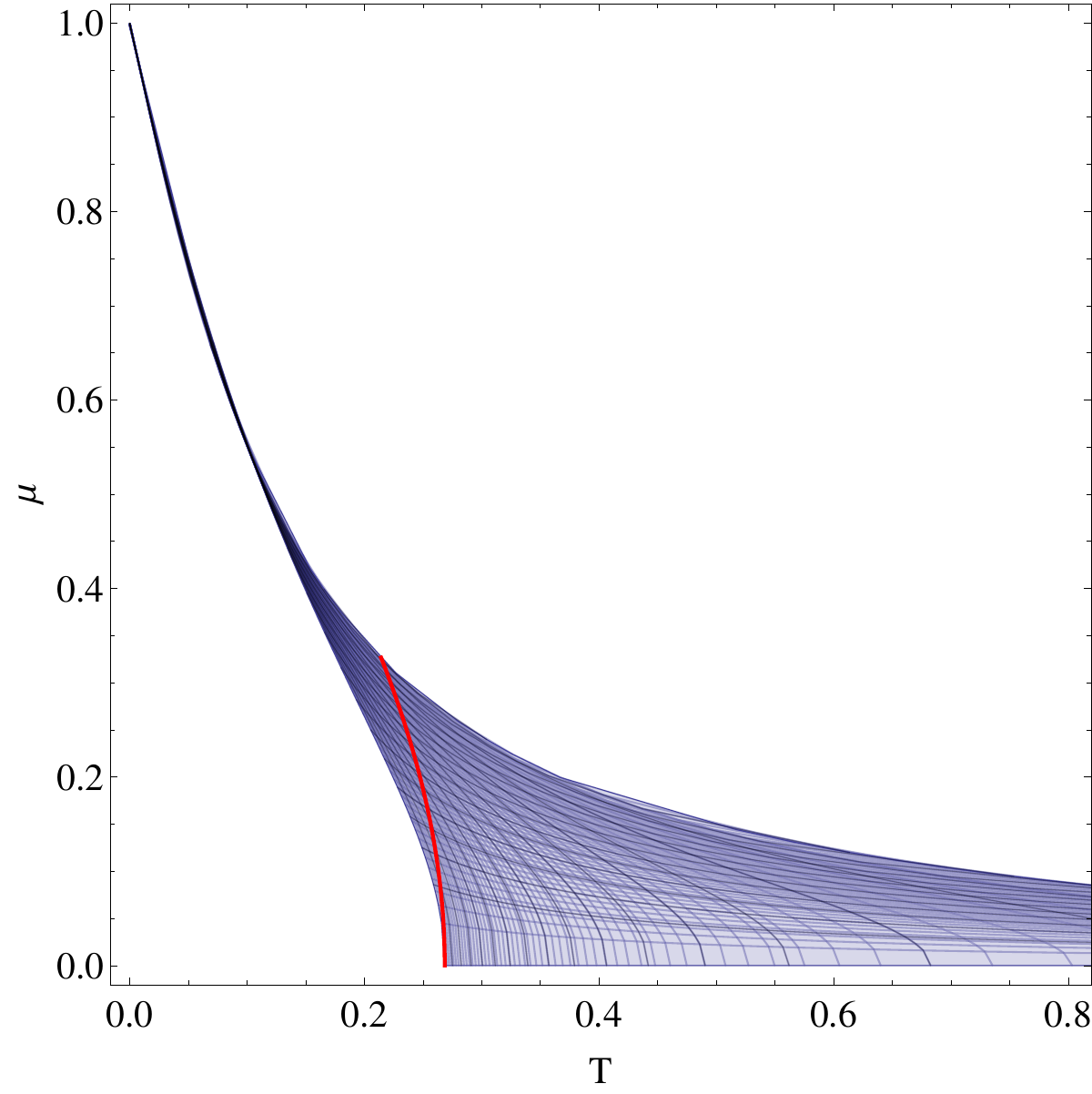}
\caption{Blue region indicates the $(T,\mu)$ values for which large black hole solutions exist.}
\label{verify}
\end{figure}

The reason this will be a phase boundary may not be intuitive, therefore, let us look at it in more detail. We can invert the relations of $T$ and $\mu$ in terms of $\rh, Q$, which can be solved only numerically. Using this we can verify that for a given $(T,\mu) $ there could be upto three solutions, of which at most only one could have $\rh>\frac{8}{9}\rb $. This means there are values of $(T,\mu) $ for which there are no solutions that correspond to large black holes (with $\rh>\frac{8}{9}\rb $). In Fig.\ref{verify}, we have the region in the $(T,\mu)$-plane which has a large black hole solution marked in blue. The Hawking-Page like curve is plotted in red, which, as one can see, falls and ends within the region marked in blue. The upper boundary of the blue region is the red curve marked in Fig.\ref{rnphasedig}.

\section{Turning on the Scalar: Hairy Solutions}

We will now add a charged scalar to this system. The result of \cite{NoHair} shows that stationary charged black hole in asymptotically flat space is completely characterized by the mass, angular momentum and charges (of the Maxwell fields), this is the \emph{no-hair theorem}. This means that asymptotically flat spaces will not support any non-trivial scalar profile. To be more concrete, let us add a scalar piece to the action \eqref{action},
\bea\label{scaction}
S_{scalar} = -\dfrac{1}{16\pi}\int_{\man}d^{4}x \sqrt{-g}\; |\nabla\psi - i q A \psi|^{2}.
\eea
The metric and gauge field have the same functional forms as that without the scalar. Using the fact that the $r$-component of Maxwell field equation forces the phase of the scalar to be a constant, which can then be absorbed by a gauge transformation (see eg. \cite{HHH1, BKA}), we take the scalar to be real, $\psi=\psi(r)$. The equations of motion\footnote{During the refereeing process of our paper we have checked that these equations are equivalent to Eq. (2.15) in \cite{Win1}, as can be seen by mapping our variables $ \{g, \psi, q, \phi\} $ onto $ \{f, \phi, \sqrt{2} q, A_0/\sqrt{2}\} $ in \cite{Win1}.} for this choice of the fields is given by
\bea
\frac{1}{2} \psi '(r)^2+ \frac{g'(r)}{r g(r)}+\frac{q^2 \psi (r)^2 \phi (r)^2}{2 g(r)^2 h(r)}+\frac{\phi '(r)^2}{g(r) h(r)}-\frac{1}{r^2 g(r)}+\frac{1}{r^2} = 0,\\
 h'(r) -r h(r) \psi '(r)^2  -\frac{r \, q^2 \psi (r)^2 \phi (r)^2}{g(r)^2 }= 0,\\
\phi ''(r)+\frac{2 \phi '(r)}{r}-\frac{h'(r) \phi '(r)}{2 h(r)}  -\frac{q^2 \psi (r)^2 \phi (r)}{2 g(r)} = 0,\\
\psi ''(r)+\frac{g'(r) \psi '(r)}{g(r)}+\frac{h'(r) \psi '(r)}{2 h(r)}+ \frac{2 \psi '(r)}{r} +\frac{q^2 \psi (r) \phi (r)^2}{g(r)^2 h(r)} = 0.
\eea 
If we want look at asymptotically flat space solution, we can expand the fields $g, h, \phi$ and $\psi$ around $r\rightarrow\infty$ in powers of $1/r$. Now plugging these solutions back into the equations of motion, and solving the equations order by order, we will get that all the coefficients in the expansion for $\psi$ will be forced to zero, and we will end up with RN-Black hole as the general solution. This is the way in which the no-hair theorem manifests itself in our set up.

But if the manifold has a boundary at $r=\rb $, we can again perform a series expansion of the four fields around $r=\rb $ and plug it back into the equations of motion, and solve the coefficients order by order. This gives the boundary functions in terms of $\psi_{0}^{b}=\psi(\rb),\psi_{1}^{b}=\psi'(\rb),\phi_{0}^{b}=\phi(\rb),\phi_{1}^{b}=\phi'(\rb),g_{0}^{b} = g(\rb),h_{0}^{b}=h(\rb)$, as
\bea
&\psi(r) = \psi _0{}^b+\left(r-r_b\right) \psi _1{}^b+ \dots ,\\
&\phi(r) = \phi _0{}^b+\left(r-r_b\right) \phi _1{}^b+ \dots ,\\
&g(r) = g_0{}^b+\left(r-r_b\right) \left(\frac{1-g_0{}^b}{r_b}-\frac{r_b \left( \left(g_0{}^b\right){}^2 h_0{}^b \left(\psi _1{}^b\right){}^2+2 g_0{}^b \left(\phi _1{}^b\right){}^2+ q^2 \left(\psi _0{}^b\right){}^2 \left(\phi _0{}^b\right){}^2\right)}{2 g_0{}^b h_0{}^b}\right)+ \dots ,\\
&h(r) =h_0{}^b+\left(r-r_b\right) r_b \left(\frac{q^2 \left(\psi _0{}^b\right){}^2 \left(\phi _0{}^b\right){}^2}{\left(g_0{}^b\right){}^2}+h_0{}^b \left(\psi _1{}^b\right){}^2\right)+ \dots .
\eea
The expansions at $r = 0$ or $r= \rh $ for the boson star and hairy black hole respectively, along with the boundary conditions, are discussed when we look at the specific solutions.

At this point, it seems relevant to discuss the some aspects of the scalar field. For the Einstein-Maxwell system, the information contained in a box is essentially the same as that in the asymptotic case. However, this is not the case for the scalar field. Taking the limit $\rb\rightarrow\infty$ is subtle, as the asymptotic space cannot support the scalar hair. In evaluating the free energy, this manifests as the Brown-York quasilocal energy definition being insufficient to capture the mass of the scalar. We will not try to propose an alternate definition for the quasilocal energy , instead we will evaluate the free energy using the on-shell action, $F = -T \log\mathcal{Z} = T S_{cl}$. We will discuss these points further in the conclusions.

We will now explicitly construct hairy solutions. There are two such classes of solutions, those without horizons and those with horizons. The former will be called a boson star (in analogy with similar solutions in AdS) and the latter is the hairy black hole.

\subsection{Boson Star}
The boson star is a a horizon-less configuration. At $r=0$, the derivatives of all the functions are set to zero. At $r=\rb$ we set Dirichlet boundary condition for the scalar, $\psibnd=0$. The expansions of the functions around $r=0$, such that they solve the equations of motion, are calculated to be
\bea
\psi(r) = \psizero - \dfrac{q^2 \phizero{}^2 \psizero  }{6 \hzero}r^2 + \dots ,\\
\phi(r) = \phizero + \dfrac{1}{12} q^2 \phizero \psizero{}^2  r^2 + \dots , \\
g(r) = 1 - \dfrac{q^2 \phizero^2 \psizero{}^2 }{6 \hzero}r^2 + \dots ,\\
h(r) = \hzero + \dfrac{1}{2} q^2 \phizero^2 \psizero^2  r^2 +\dots .
\eea
Here, we have $\psizero =\psi(0)$, $\phizero = \phi(0)$ and $\hzero = h(0)$, which parametrize the solutions, and all the six boundary parameters are determined from these three. The value of $\hzero$ can be arbitrary as we have to rescale the function $h(r)$ at the end to have the right boundary behaviour. The solutions are found by fixing a value for $\psizero$, setting $\hzero =\, $const., say 1, and choosing $\phizero$ such that $\psibnd$ is zero. 

The boson star configuration can have arbitrary temperature, and the value of chemical potential above which it can exist is controlled by $q$. This point of instability of the flat empty box to forming a boson star can be calculated analytically. At the point of instability, the scalar profile is not strong enough to cause any backreaction. Thus, we can take $\psi(r)\rightarrow \alpha\psi(r)$, where $\alpha\ll 1$, and look at the equations of motion upto linear order in $\alpha$. With the given boundary conditions, we will get the solution $g(r)=1, h(r)=1, \phi(r)=\mu $, and the scalar equation of motion gives
\bea
\psi''(r) + \dfrac{2}{r}\psi'(r) + \mu^{2}q^{2} \psi (r) =0.
\eea
Imposing the scalar boundary conditions, we get the solution
\bea\label{bsi}
\psi(r) = \psizero \dfrac{\sin \mu q r}{r}, \ \text{with} \ \mu_{bsi} q = n \pi, \ n= 1,2,3,\dots \;.
\eea
We will be looking at the first eigenmode, i.e. $n=1$. 

\begin{figure}
    \centering
    \begin{subfigure}[b]{0.4\textwidth}
        \includegraphics[width=\textwidth]{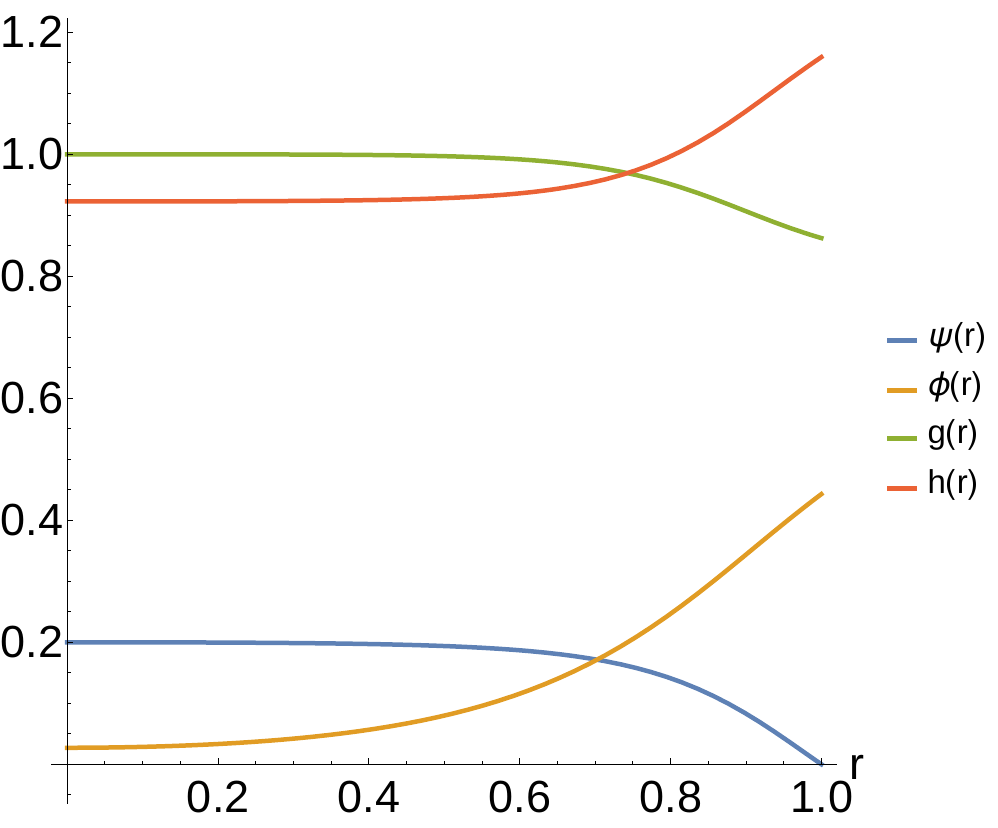}
        \caption{$q=40,\psi_{0} =0.2$}
         \end{subfigure}
    \begin{subfigure}[b]{0.4\textwidth}
        \includegraphics[width=\textwidth]{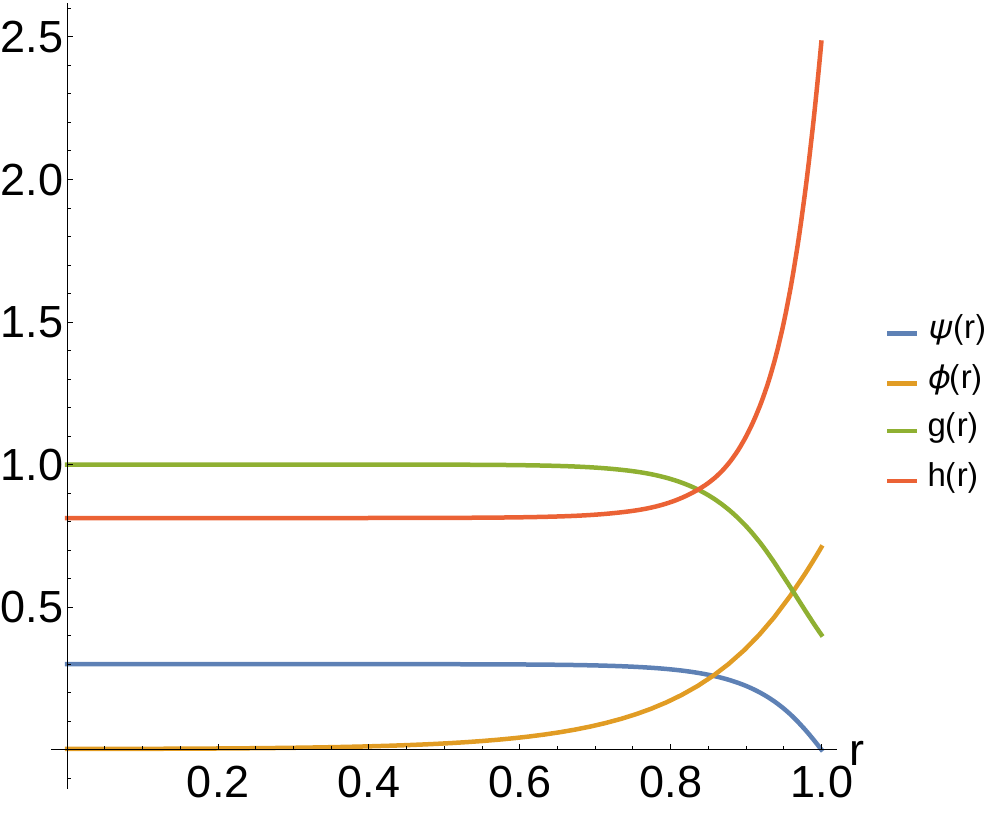}
        \caption{$q=40,\psi_{0} =0.3$}
         \end{subfigure}
     \caption{Sample profiles of $\psi(r), \phi(r),g(r)$ and $h(r)$ in the fully backreacted solution for boson star, with $\rb=1$. The quantities on the y-axis are labelled to the right of the figure.}\label{bsplot}
\end{figure}

Of course, we can also construct fully backreacted solutions as well, numerically. In Fig.\ref{bsplot}, we have shown the profiles of the functions for two fully backreacted solutions.

\subsection{Hairy Black Hole}

The hairy black hole is a system with a horizon and a non-trivial scalar profile. The existence of solutions in the box shows that the no-hair theorems of the asymptotic space do not apply when one is looking at a box. The boundary conditions at the horizon are $g(\rh)=0$ and $\phi(\rh)=0$, the latter ensures that the Maxwell field is regular at the horizon. Around $r=\rh $, the functions can be written as a series, such that they solve the equations of motion,
\bea
&\psi(r) = \psiin - \dfrac{ q^2  \rh^2 \hin \phiin^2 \psiin }{
 4 (\hin - \rh^2 \phiin^2 )^2} (r - \rh)^2 +\dots , \\
&\phi(r) = \phiin (r - \rh) + \dfrac{ \phiin (8\rh^2 \hin \phiin^2  - 4\rh^4 \phiin^4  +  \hin^2 ( q^2 \rh^2 \psiin^2 -4 ))}{4 \rh (\hin - \phiin^2 \rh^2)^2}(r - \rh)^2  + \dots , \\
&g(r) =  (\dfrac{1}{\rh} - \dfrac{\phiin^2 \rh}{\hin}) (r - \rh)- \dfrac{4 \hin^2 + 8 \rh^4 \phiin^4 \rh^4 + 3\rh^2 \hin \phiin^2  (q^2 \rh^2 \psiin^2  -4 )}{ 4  \rh^2 \hin (\hin - \rh^2 \phiin^2 )}  (r - \rh)^2 + \dots ,  \nonumber \\
\\
&h(r) = \hin + \dfrac{\hin^2 \phiin^2 \psiin^2 q^2  \rh^3}{(\hin - 
   \phiin^2 \rh^2)^2}  (r - \rh)  
   + \dots ,
\eea
where $\psiin= \psi(\rh)$, $\phiin= \phi'(\rh)$ and $\hin= h(\rh)$. The value of the six parameters at the boundary are determined from the choice of these three parameters. The choice of $\hin$ is arbitrary as the solution is rescaled at the end to get the correct boundary metric. Thus, we set $\hin=1$, and tune $\phiin$ such that $\psibnd =0 $, for different values of $\psiin$, $q$ and $\rh$, and then appropriately rescale the functions $\phi$ and $h$.

\begin{figure}[H]
    \centering
    \begin{subfigure}[b]{0.4\textwidth}
        \includegraphics[width=\textwidth]{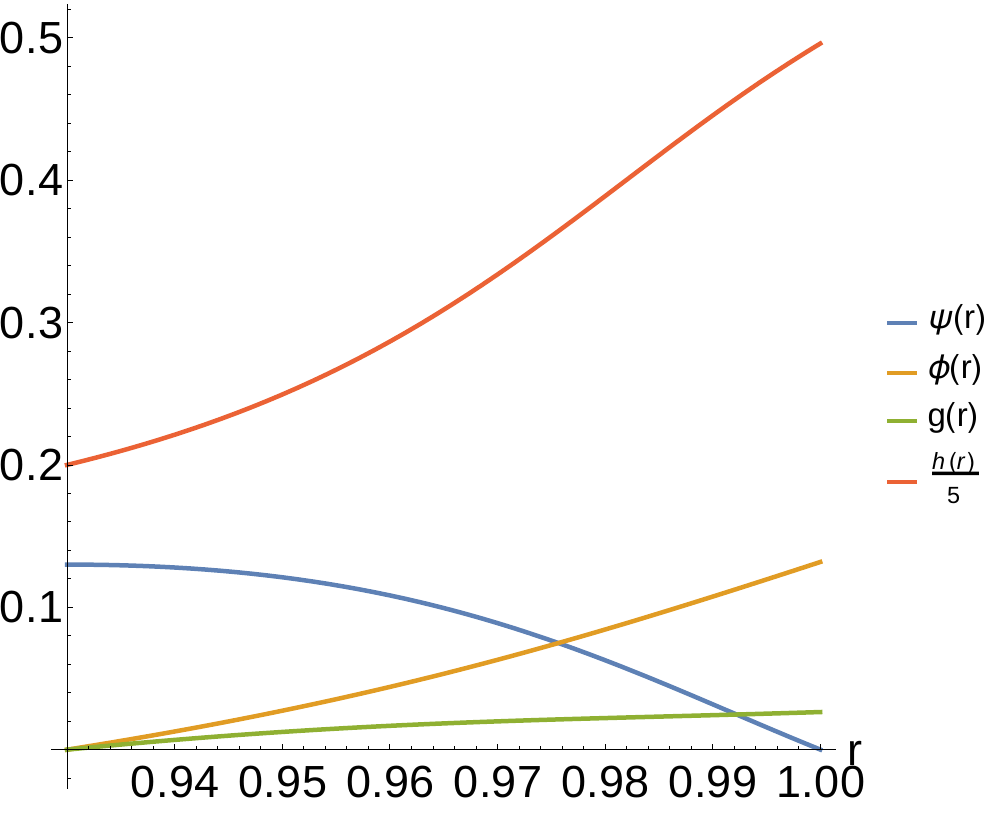}
        \caption{$q=30,\rh = 0.93, \psi_{0} =0.13$}
         \end{subfigure}
    \begin{subfigure}[b]{0.4\textwidth}
        \includegraphics[width=\textwidth]{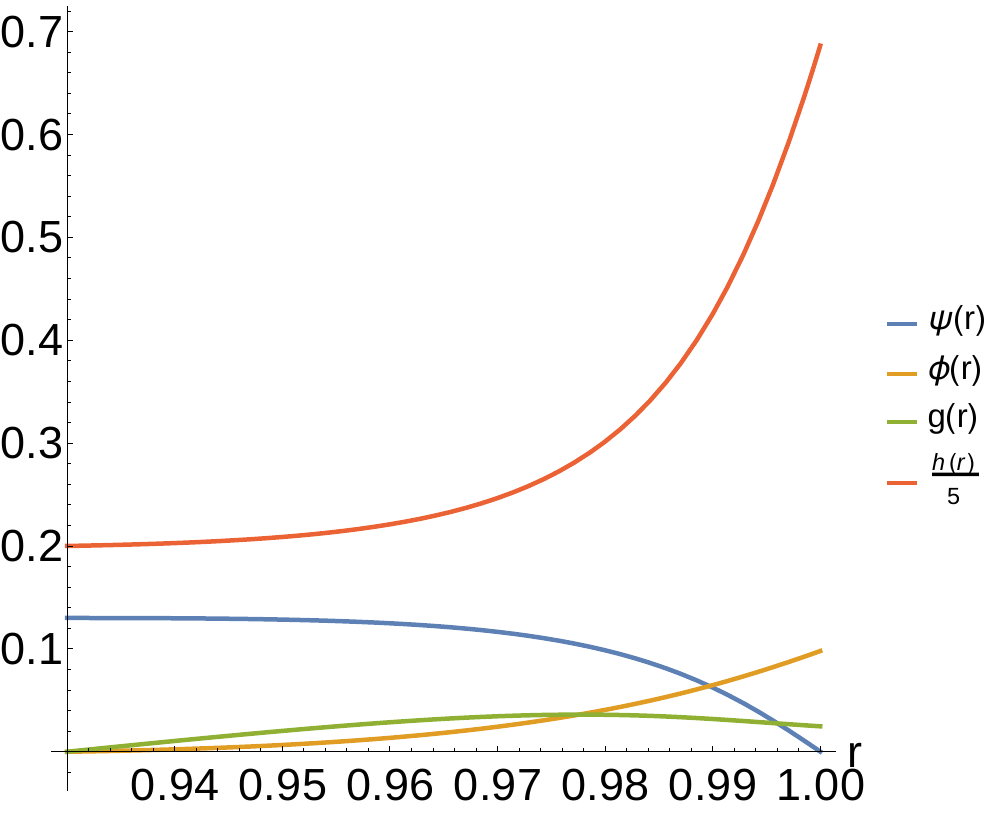}
        \caption{$q=100,\rh=0.93 ,\psi_{0} =0.13$}
         \end{subfigure}
     \caption{Sample profiles of $\psi(r), \phi(r)$, $g(r)$ and $\frac{h(r)}{5}$ in the fully backreacted solution for hairy black hole, with $\rb=1$. The quantities on the y-axis are labelled to the right of the figure.} \label{bhprofiles}
\end{figure}

The instability of a RN black hole to develop hair is dependent on $q$ and $\rh$, and cannot be evaluated analytically. In the limit $\psi\rightarrow \alpha\psi(r)$ with $\alpha\ll 1$, and looking at upto terms linear in $\alpha$, we get the RN solution, and a homogeneous equation for $\psi(r)$, which can be solved numerically to find the first eigenmode,
\bea
g(r)= 1 -\dfrac{1}{r}\left(\rh+ \dfrac{Q^{2}}{\rh }\right), \ h(r) = \dfrac{1}{g(\rb)}, \ \text{and} \ \phi(r) = \dfrac{Q}{\sqrt{g(\rb)}} \left(\dfrac{1}{\rh} - \dfrac{1}{\rb }\right),\\
\psi ''(r) +\dfrac{\left(Q^2-2  \rh\,r+\rh^2\right)}{(r-\rh) \left(Q^2-\rh \,r \right)}  \psi '(r) + \dfrac{q^2 Q^2 r^2 }{\left(Q^2-\rh r\right)^2}\psi (r)= 0.
\eea

The profiles of two fully backreacted solutions are given in Fig.\ref{bhprofiles}.

\section{The Phase Diagram}

As we discussed earlier, the free energy of the system when there is a non-trivial scalar profile present is done by evaluating the on-shell action. The full action is given by the sum of \eqref{action} and \eqref{scaction}. Since the boundary metric of all the systems are rescaled to be of the same form as the boundary metric of empty box, the temperatures of all the systems can be consistently compared. Using the equations of motions, we can rewrite the action (for details of a similar calculation see Appendix of \cite{BKS1602})
\bea
F = \dfrac{S_{euc}}{\beta} =-\dfrac{\int_{r_0}^{\rb}\sqrt{h(r)}dr}{2\sqrt{g(\rb)h(\rb)}}-\dfrac{1}{2}\rb\sqrt{g(\rb)}- \dfrac{\rb^{2} \left(g'(\rb)h(\rb) +h'(\rb)g(\rb) \right)}{4\sqrt{h(\rb)}} + \rb.
\eea
The phase diagram is intricately dependent on $q$, which gives three distinct types of phase diagrams, which have two, three or four of the four possible solutions as thermodynamically acceptable solutions. We will look at each of theses cases in detail.

\subsection{$q_{1}$ < $q$ < $\infty$ }
\begin{figure}[H]
\centering
\includegraphics[width=0.6\textwidth]{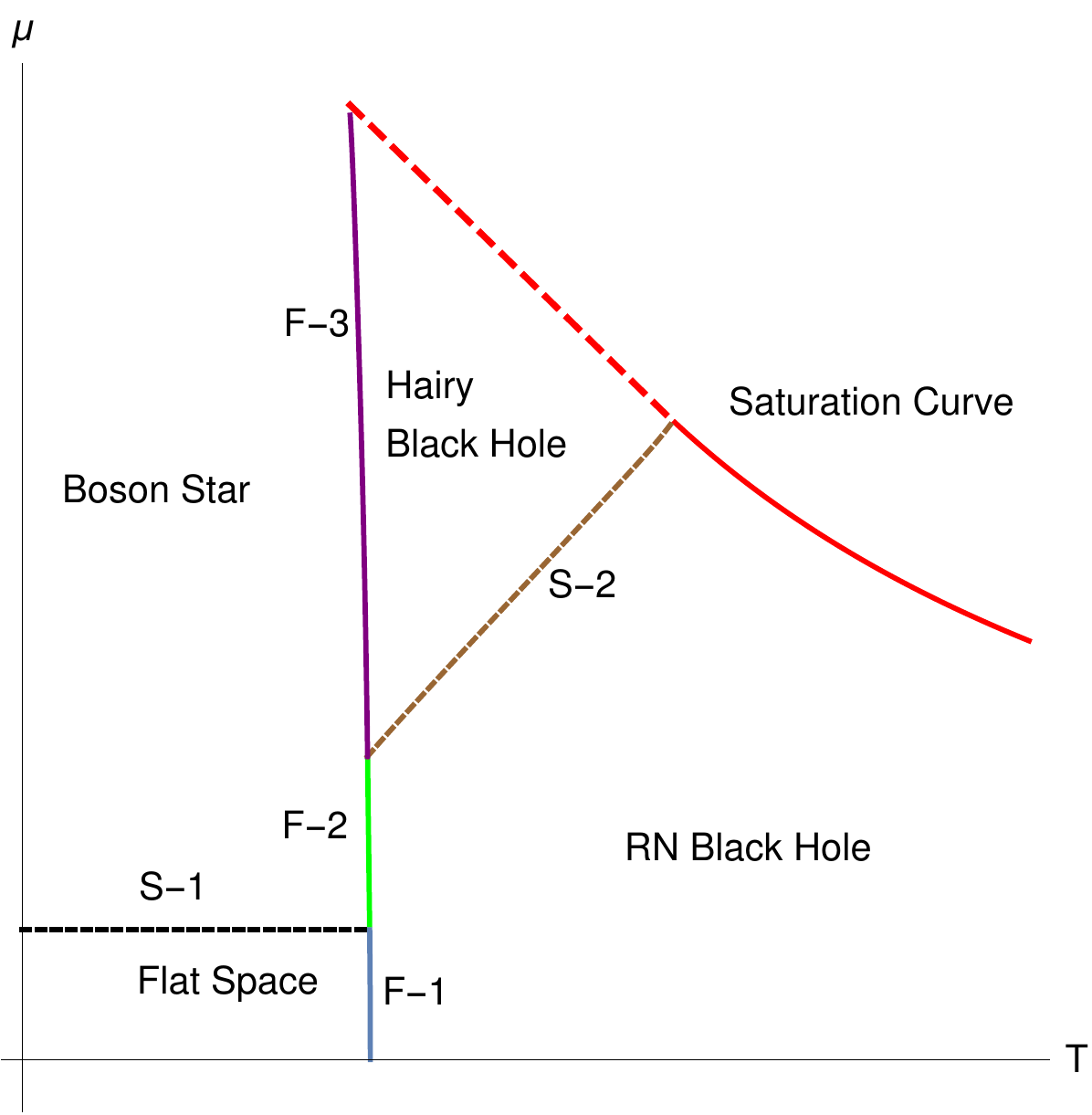}
\caption{Schematic phase diagram for $\infty<q<q_{2}$.}
\label{largeqsample}
\end{figure}
For this case all of the four solutions appear in the phase diagram in certain regions of the $ T-\mu $ plot. A representative diagram is given in Fig.\ref{largeqsample}. For values of  $\mu $ smaller than the boson star instability, $\mu_{bsi}$, for the given value of $q$ given by \ref{bsi}, the phase boundary, $\mathit{F}$-1, separates the empty flat box from the RN black hole, which is a first order phase transition, and can be computed analytically. Above the value at which the boson star instability happens, say $\mu_{bsi}$, the favourable phase is a boson star, which is a second order phase transition from the flat empty box, indicated by $\mathit{S}$-1, and has lower free energy than the empty box. The phase transition between boson star and RN black hole is first order in nature and th phase boundary is indicated by$\mathit{F}$-2, which is computed semi-analytically by the following method. For a given value of $q$, the value of $\mu $ and $F$ can be determined for each value of $\psizero$, and we can do a fit to get free energy as a function of $\mu$, and then find the value of $T$ for the RN black hole with the same value of $\mu$ and $F$.

The curve $\mathit{F}$-2 comes to an end at the point where it intersects the hairy black hole instability curve, $\mathit{S}$-2. The RN black hole to hairy black hole transition is also a second order phase transition. One can check that for a given value of $T$ and $\mu$ (where the hairy black hole solution exists), the hairy black hole has a lower free energy than a RN black hole with the same $(T,\mu)$. The phase boundary between the boson star and hairy black hole is another first order transition, which is given by the curve $\mathit{F}$-3. This curve is slightly more difficult as the values of $(T,\mu,F)$ for both the competing phases are found by {\em numerically} solving the fully backreacted equations of motion of the respective systems. For the boson star case, we have the free energy as a function of $\mu$. In the case of the box, for a hairy black hole of a larger value of $\rh $ than the one of the black hole residing at the intersection of $\mathit{F}$-2 and $\mathit{S}$-2, say $r_{h}^{c}$ will go to a boson star phase as we increase $\mu$ past some critical point (as opposed to the case in global AdS, see \cite{BKS1602}). We start with a value of $\rh >\rh^{c}$, and keep increasing the value of $\psiin$ till the free energy of the hairy black hole becomes equal to the free energy of the boson star with the same $\mu $.
\begin{figure}[H]
    \centering
    \begin{subfigure}[b]{0.4\textwidth}
        \includegraphics[width=\textwidth]{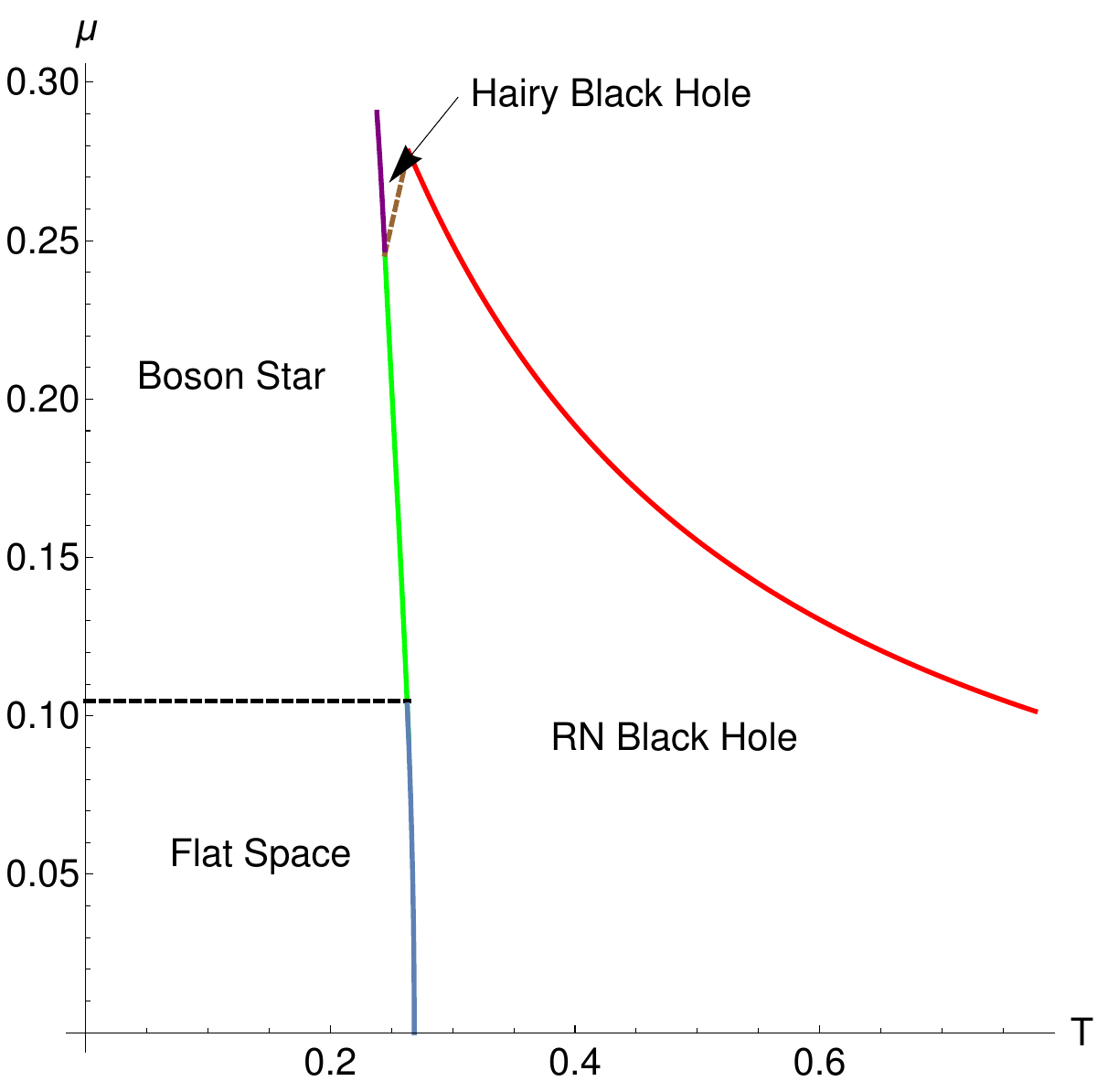}
        \caption{$q=30$}
         \end{subfigure}
    \begin{subfigure}[b]{0.4\textwidth}
        \includegraphics[width=\textwidth]{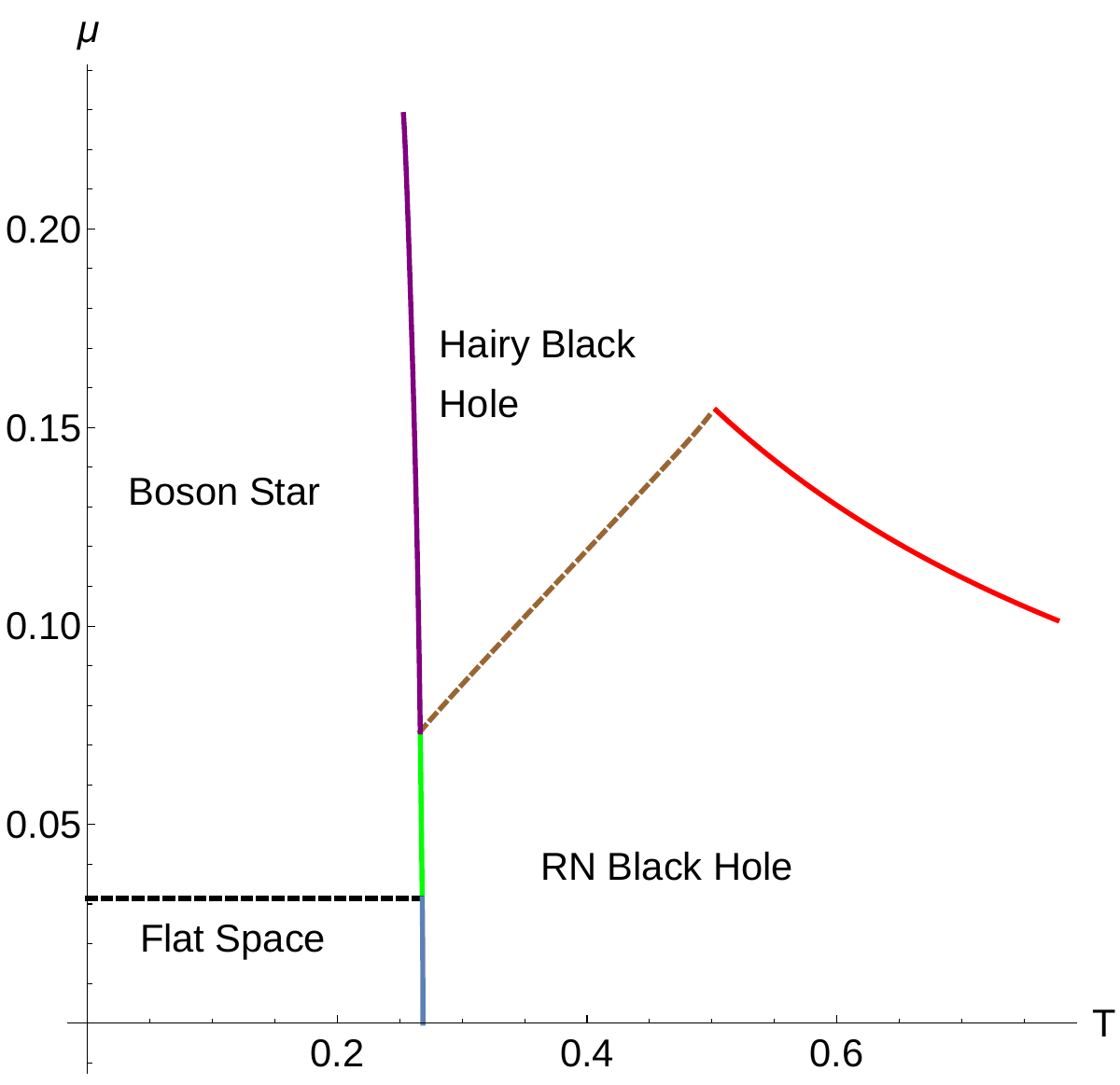}
        \caption{$q=100$}
         \end{subfigure}
     \caption{Phase diagrams for $q=30,100$.}\label{largeq}
\end{figure}
In principle the curve $\mathit{F}$-3 should be computable for all the values of $\rh$ till it becomes very close to $\rb $. However, the shooting procedure that we use to find the solutions becomes increasingly difficult as we take $\rh $ closer to $\rb$, depending on the value of $q $. For example, for $q=100$, we are able to obtain the curve upto around $\rh=0.9075$, and for $q=30$, we can go to about $\rh=0.9575$ (all with $\rb =1$). Near these numerical limits, what we find is that the curve $ \mathit{F}-3 $ is very close to where the saturation curve of the RN Black Hole would be for the hairless solutions. In other words, the exact structure of the red-dotted line in our schematic diagram Fig. \ref{largeqsample} cannot be precisely obtained with our current numerics. That the hairy black hole region has to be bounded is based on the fact that the black hole size is limited by the box. The precise form of the way the region closes as $\rh\rightarrow\rb $ does not alter the punchline that the hairy black hole is a thermodynamically favourable phase in some regions of the $(T,\mu)$-plane. So we will relegate that to future work.

The exact phase diagrams for $q=30,100$ are shown in Fig.\ref{largeq}. As we can see that region in which the hairy black hole can exist shrinks as we go to smaller values of $q$. Eventually, we end up with the case where $\rh^{c}\rightarrow\rb $, which happens at $q=q_{1}$. It is difficult to find the exact value of $q_{1}$ as the numerical value will have to be found by a tedious trial and error method. However, we have found that it happens for $q \approx 24 $.

\subsection{$q_{2}<q<q_{1}$}
\begin{figure}[H]
\centering
\includegraphics[width=0.6\textwidth]{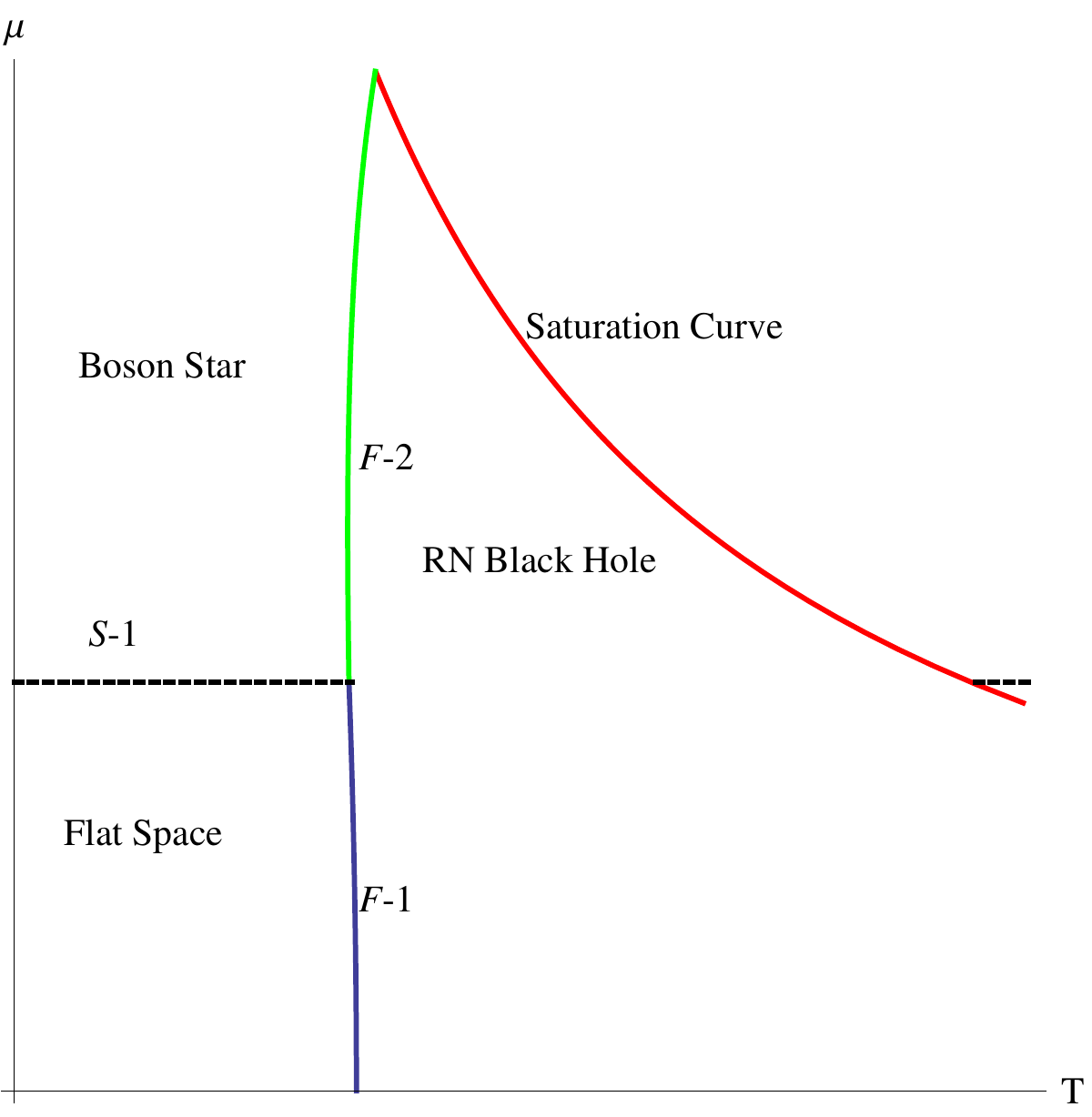}
\caption{Schematic phase diagram for $q_{2}<q<q_{1}$.}
\label{intersample}
\end{figure}
As we can see from the discussion in the previous case, this range of $q$ will give a phase diagram where only three of the four solutions can exist as thermodynamically favourable phases, namely flat empty box, RN black hole and boson star, see Fig.\ref{intersample}. The curves $\mathit{F}$-1, $\mathit{S}$-1 and $\mathit{F}$-2 are computed using the same procedure as mentioned for the large $q$ case. The RN black hole instability happens for values of $(T,\mu)$ where the RN black hole itself is not the favourable phase, whereby the second order transition does not happen, and the hairy black hole does not appear as a thermodynamically favourable phase. As a result, the curve $\mathit{F}$-2 ends when it intersects with the saturation curve.

\begin{figure}[H]
    \centering
    \begin{subfigure}[b]{0.4\textwidth}
        \includegraphics[width=\textwidth]{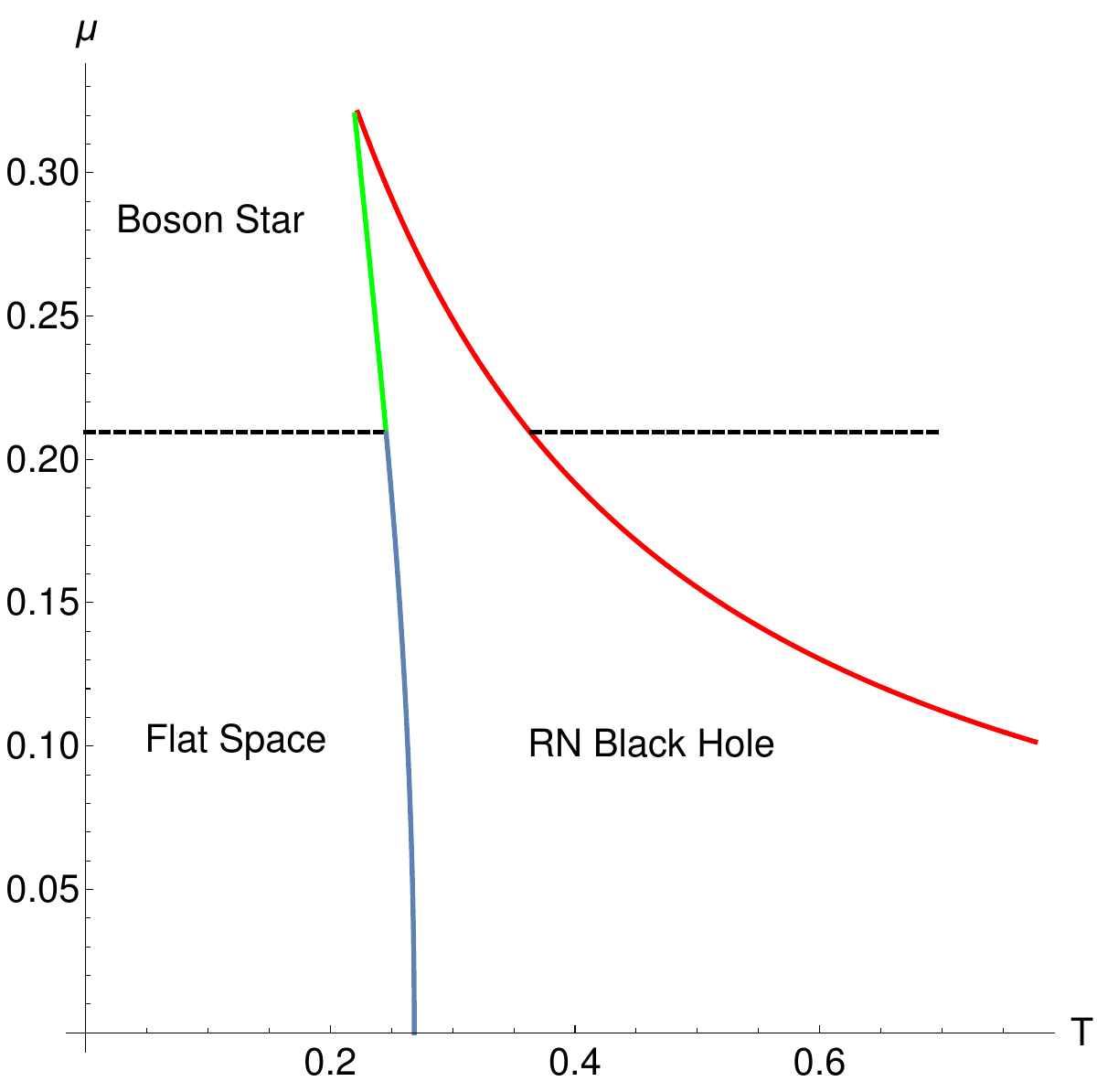}
        \caption{$q=15$}
         \end{subfigure}
    \begin{subfigure}[b]{0.4\textwidth}
        \includegraphics[width=\textwidth]{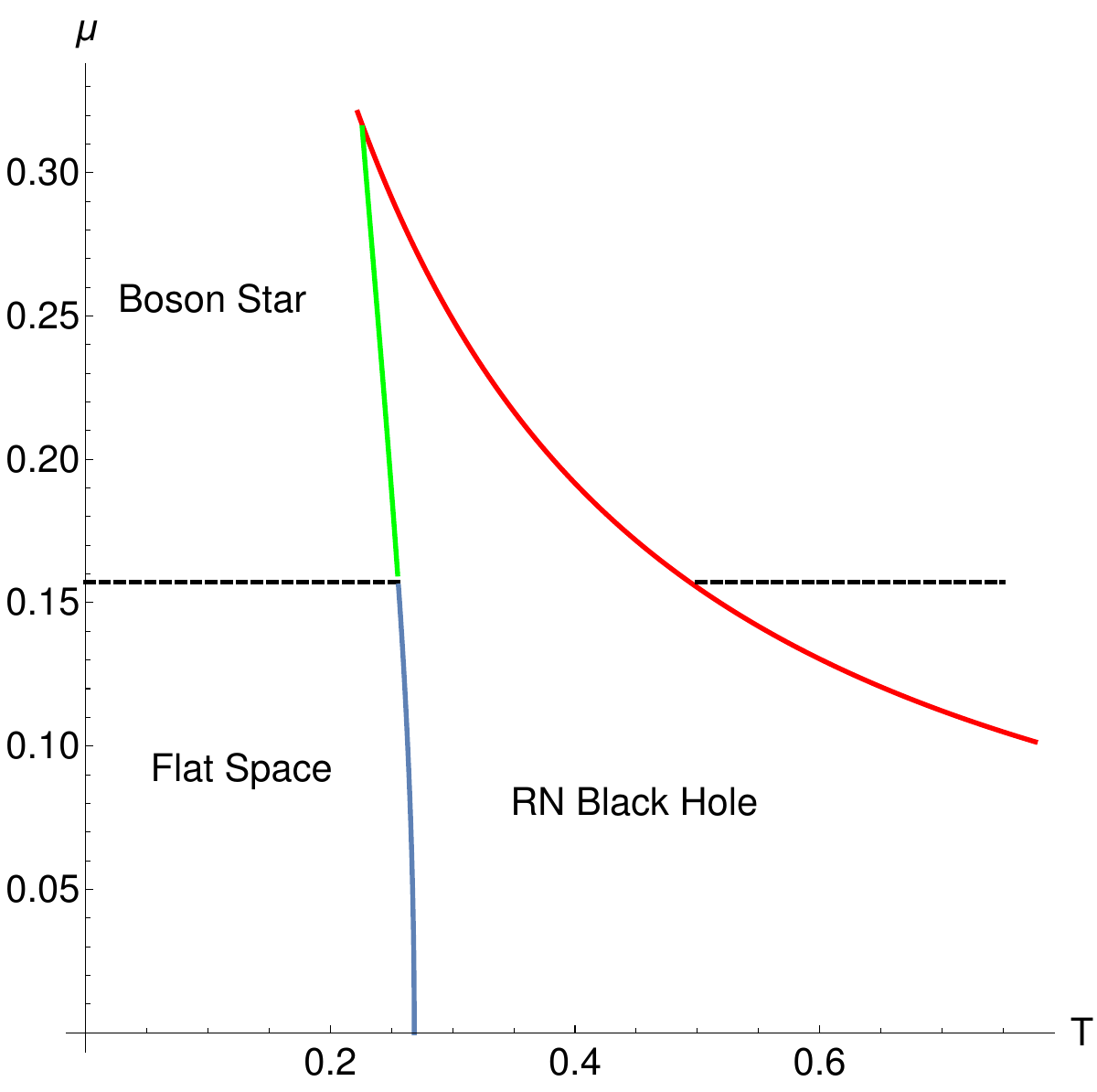}
        \caption{$q=20$}
         \end{subfigure}
     \caption{Phase diagrams for $q=15,20$.}\label{interq}
\end{figure}

We have plotted the exact phase diagram for $q=15,20$ in Fig.\ref{interq}. This type of phase diagram can exist only until the values of $q$, such that the value of $\mu_{bsi}<\frac{1}{3}$, or in other words $q<q_{2} = 3\pi$, below which there will be no phase boundary between RN black hole and boson star.

\subsection{$q_{3}<q<q_{2}$}

\begin{figure}[H]
\centering
\includegraphics[width=0.6\textwidth]{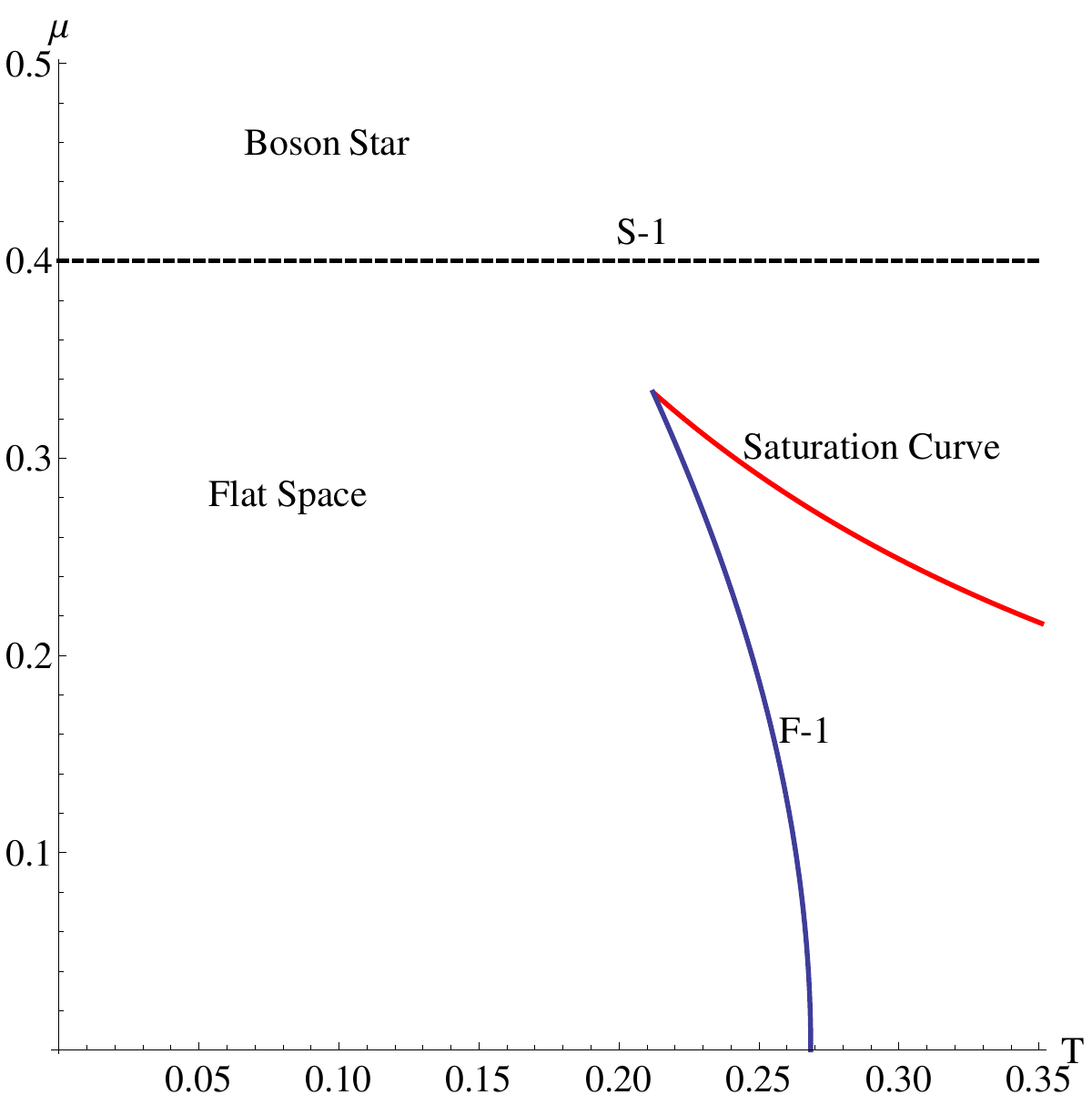}
\caption{Phase diagram for $q=7.85$.}
\label{smallsample}
\end{figure}

For values of $q_{3}<q<q_{2} = 3\pi$, the boson star instability happens at $\mu_{bsi}>\frac{1}{3}$, which in some sense leads to a trivial extention of the phase diagram of scalar-less case, see Fig.\ref{smallsample}. The boson star becomes the dominant phase above $\mu=\frac{\pi}{q}$. 
This sort of phase diagrams, where the boson star exist as a stable phase only occurs till $ q = q_{3}\approx 2 $.

\subsection{$ q< q_{3} $}
For $ q<q_{3}\approx 2 $, the fully backreacted boson stars behave very differently to $ q>q_{3} $. For $ q<q_{3} $, the chemical potential of the boson star configuration reduces from the corresponding $ \mu_{bsi} $ as we increase the backreaction by adding larger and larger scalar profiles, unlike in any of the cases seen for $ q>q_{3} $. We can also verify that for such configurations, the free energy of the solution is greater than the flat space solution, which renders them thermodynamically unstable as well. (See \cite{New}, for more details on a similar behavior in $ 2+1 $-d)

\section{Conclusions, Discussion and Future Directions}

We have charted out the phase diagram of the Einstein-Maxwell-scalar system in a box and demonstrated the existence of thermodynamically stable hairy solutions with and without horizons. The results that we find have close analogues in global AdS, but also some differences. These differences are closely tied to the fact that the box is a hard cut-off on the size of the black hole: extremal limit of thermodynamically stable black holes in AdS correspond to zero size black holes (both inner and outer horizon shrink to zero size), whereas in the box they correspond to the outer and inner horizon reaching the box size simultaneously. Our phase diagram is nearly complete, but it will be nice to precisely chart the upper boundary of the hairy black hole solutions where the black hole size reaches that of the box. We also find that there is a certain cut-off on $ q $ below which the transition to boson star from flat space is not thermodynamically favourable. An entirely analogous phenomenon also occurs in 2+1 dimensional case, as elaborated in \cite{New}. Note also that the same phenomenon happens in AdS as well \cite{BKS1602}, but the mechanism is different. There, as is clear from (say) Figure 1 of \cite{BKS1602}, there is an upper bound for $\mu$ where empty global-AdS can exist and it is this, together with the (perturbative) boson star instability condition $q \mu=1$, that makes it plausible that $q$ has a lower bound. Here on the other hand, Figure 6  does not lead to an automatic bound on $\mu$, but the free energy ends up resulting in a bound on $q$ anyway. The fact that AdS and the box match up in this subtle way, is satisfying.

One interesting observation is that having a non-trivial scalar in the box means that in the limit where the box size goes to infinity, these solutions are {\em not} the within-box truncation of the asymptotically flat (and therefore hairless) solutions. One might wonder what makes the scalar case and the gauge field so different when they are considered in a box. The point is that holding the gauge field fixed at a finite radius contains essentially the same information as fixing it at infinity (once the $r_h$ and $r_b$ are given). There exists a simple limit for gauge field when the $r_b \rightarrow \infty$, where the gauge field remains non-vanishing. But this is {\em not} the case for scalar. The $r_b \rightarrow \infty$ limit is non-trivial for the scalar, because no hair theorems force the scalar to be zero at all orders in $1/r$ in that limit.

This has consequences for defining the quasi-local mass in a box in the sense of \cite{BrownYork}. Typically, the quasi-local mass just puts a box around some region to define the mass/charges in that regions {\em while} allowing the field itself to decay to its asymptotically flat values outside the box. But when we have a non-trivial scalar profile in the box the quasi-local definition of mass will not work because it is not just about putting a box around the region of interest, but also about changing the boundary conditions of the scalar at the box. (This is not the case for the gauge field, where non-trivial boundary conditions at the box are automatically obtained by fixing them at infinity.) That the quasi-local mass cannot work is straightforward to check, because thermodynamic relations of the form $F \sim E-TS -\mu Q$ etc do not work in the box, if we use the quasi-local definitions. Fortunately, we do not need explicit forms of these quantities to chart out the phase diagram, we just need the ability to compute the action aka free energy directly. That, together with the fact that the phase diagram has internal consistency (the various independent curves in it intersect consistently and the overall structure matches very closely with that of AdS \cite{BKS1602}) give us confidence that the results are correct. 

It should be possible to generalize the definitions of quasi-local quantities so that one can define a thermodynamically useful notion of the mass of a scalar with a non-trivial profile in a finite region. We hope to come back to this interesting problem in the future. It seems evident that the box boundary acts a compression cavity to hold the scalar in, and therefore a pressure-like term will have to be added to the total mass of the spacetime.

It will also be interesting to consider extremal solutions (not necessarily thermodynamically stable) and perhaps see the possibility of attractor behavior. Note however that attractor behavior is typically associated to uncharged scalars, so the flavor here is slightly different. One can also try to construct hairy solutions with other boundary conditions (see eg., \cite{Neumann,Neumann1,Neumann2}). One other interesting line is to consider generalizations of these solutions to higher dimensions: there exists a large class of solutions in higher dimensions \cite{5d} with a rich phase structure (see also \cite{Lahiri} for a ``dual" analysis of phases) and it will be interesting to see how adding a scalar (in the box) changes these results. 

\section*{Acknowledgments}
We thank Daniel Arean, Tanmoy Das, Justin David, Carlos Herdeiro, Madhavan, Suvrat Raju, Joseph Samuel, Aninda Sinha, Sumati Surya and K. P. Yogendran for comments/discussions.

\end{document}